\documentclass[aps,prd,10pt,twocolumn,showpacs,amsmath,amssymb,nofootinbib,superscriptaddress]{revtex4-2}
\usepackage{graphicx}  
\usepackage{color}
\usepackage{times}
\usepackage{float}
\usepackage{afterpage}
\usepackage[utf8]{inputenc}
\usepackage{bm}
\usepackage{ulem}
\usepackage{multirow}
\usepackage{makecell}
\usepackage{url}
\usepackage{natbib}
\usepackage{mathrsfs}
\usepackage{physics}
\usepackage{comment}
\usepackage[table,xcdraw]{xcolor}
\usepackage{booktabs,makecell}
\usepackage{amsmath}
\usepackage{pifont}
\usepackage[colorlinks=true,citecolor=blue,urlcolor=blue,linkcolor=blue]{hyperref}
\usepackage{microtype}
\usepackage[none]{hyphenat}
\begin{document}

% Title and author information
\title{Slowly Rotating Two-Fluid Neutron Stars: Coupled Frame-Dragging, Inertia Splitting, and Universal Relations}

% Authors and affiliations
\author{Ankit Kumar}
\email{ankitlatiyan25@gmail.com}
\affiliation{Department of Mathematics and Physics, Kochi University, Kochi, 780-8520, Japan}
\affiliation{Department of Physics, Indian Institute of Science, Bangalore 560012, India}

\author{Hajime Sotani}
\affiliation{Department of Mathematics and Physics, Kochi University, Kochi, 780-8520, Japan}
\affiliation{RIKEN Center for Interdisciplinary Theoretical and Mathematical Sciences (iTHEMS), RIKEN, Wako 351-0198, Japan}
\affiliation{Theoretical Astrophysics, IAAT, University of T\"{u}bingen, 72076 T\"{u}bingen, Germany}

\date{\today}
%%%%%%%%%%%%%%%%%%%%%%%%%%%%%
% Abstract
\begin{abstract}
We develop a fully relativistic framework to study the rotational response of gravitationally coupled two-fluid neutron stars within the slow-rotation approximation. Treating the two components as independently conserved perfect fluids interacting only through spacetime curvature, we derive the coupled equilibrium and frame-dragging equations and exploit their linear structure to construct a basis decomposition of the rotational response. This formulation leads to a natural definition of the effective total moment of inertia, which generalizes the single-fluid concept and depends solely on the equilibrium background. It further reveals that the coupled system admits two intrinsic collective rotational eigenmodes, characterized by distinct eigen-moments of inertia, even in the absence of relative rotation between the fluids. Applying this framework to neutron stars containing dark matter, we explore how the presence of an additional gravitationally bound component modifies the global rotational response and its relation to tidal deformability. Our results demonstrate that the persistence or breakdown of rotational-tidal universality in two-fluid neutron stars is governed by dark-sector microphysics rather than by the mere presence of an additional component, and establish a unified framework for interpreting rotational observables, intrinsic mode structure, and universal relations in multi-component relativistic stars.
\end{abstract}
%%%%%%%%%%%%%%%%%%%%%%%%%%%%%

% PACS numbers
%\pacs{Insert PACS numbers here}

\maketitle

% Introduction
%%%%%%%%%%%%%%%%%%%%%%%%%%%%%
\section{Introduction}
\label{sec:1}
%%%%%%%%%%%%%%%%%%%%%%%%%%%%%
Neutron stars are widely recognized as natural laboratories for probing fundamental physics under extreme conditions of density, gravity, and composition. In recent years, increasing attention has been devoted to the possibility that these compact objects might accrete or accumulate a dark matter component from their surroundings -- for instance, through scattering and capture of dark matter particles from the ambient halo over astrophysical timescales -- or may retain a dark matter fraction inherited during their formation in dark-matter‑rich environments~\cite{PhysRevD.40.3221, PhysRevD.81.123521, PhysRevD.82.063531, PhysRevLett.115.111301, PhysRevD.83.083512, PhysRevD.87.055012, Güver_2014, PhysRevD.77.043515, Nelson_2019}. This possibility has motivated a variety of theoretical frameworks that model dark-matter-admixed neutron stars, wherein dark matter coexists with ordinary nuclear matter in a gravitationally bound configuration~\cite{PhysRevD.77.023006, PhysRevD.111.043016, PhysRevD.104.063028, 10.1093/mnras/stac1013, PhysRevD.110.063001, universe11050159}. A particularly versatile approach employs a two‑fluid formalism, treating the neutron star as composed of two distinct fluids that share the same gravitational potential but may develop different radial profiles: one describing the baryonic (nuclear) component and the other representing the dark sector. This two-fluid treatment allows the components to respond differently to gravity, leading to configurations that can range from compact dark cores to extended halos, depending on the underlying microphysics and the central density for each fluid. In such models, the two fluids are typically assumed to interact only through gravity, without direct particle exchange or additional non‑gravitational couplings, making this framework especially useful for exploring the macroscopic signatures of gravitationally coupled dark matter in astrophysical settings~\cite{PhysRevD.103.043009, fnqt-sgyc, PhysRevD.105.123034, Kumar2025, Sagun_2023, kcl2-qgxh}.

The macroscopic properties of such gravitationally coupled two-fluid configurations have been investigated extensively in the static, spherically symmetric limit. In particular, previous studies have solved the coupled Tolman-Oppenheimer-Volkoff (TOV) equations for the baryonic and dark matter components and demonstrated that the presence of a dark sector can substantially modify global quantities such as the gravitational mass, radius, stability thresholds of central density, and tidal deformability of the star~\cite{Shawqi_2024, PhysRevD.111.083038, PhysRevD.97.123007, PhysRevD.110.023024}. The extent of these structural modifications depends sensitively on the microphysical properties of the dark matter, including its nature, particle mass, and self-interaction strength. Predictions from these models have been confronted with constraints from multimessenger astrophysics, providing bounds on the allowed dark matter content in such stars. Gravitational-wave observations from binary neutron star mergers such as GW170817~\cite{PhysRevLett.121.161101}, together with x-ray measurements of pulsar radii from the Neutron star Interior Composition Explorer (NICER) mission~\cite{Dittmann_2024, Vinciguerra_2024}, have been used to constrain the dark matter fraction and its self-interaction strength in such two-fluid stars by comparing theoretical mass-radius relations and tidal deformabilities with observational data, assuming that the observed neutron stars are described by dark-matter-admixed configurations~\cite{PhysRevD.111.123034, fnqt-sgyc, 10.1093/mnras/stad3658, PhysRevD.110.023013, Kumar2025}.

While most early investigations of dark-matter-admixed neutron stars focused on non-rotating configurations, rotational properties within two-fluid or related frameworks have begun to receive increasing attention in recent literature~\cite{4tkh-7hjs, PhysRevD.110.123019, Konstantinou_2024, qcl7-m5kf, shawqi2025rotatingneutronstarsdark}.
%While most of these investigations have focused on non-rotating stars, the rotational properties of dark-matter-admixed neutron stars in the two-fluid formalism remain comparatively less explored. 
In the slow-rotation approximation of general relativity, the stellar rotation induces a frame-dragging effect, leading to differential angular velocities of local inertial frames inside the star, even if the star rotates uniformly. The resulting frame-dragging function provides access to the total moment of inertia, which plays an important role in modeling pulsar spin evolution and in establishing equation-of-state-insensitive relations, the so-called universal relations, such as the I-Love-Q ~\cite{PhysRevD.88.023009}, which connect rotational and tidal properties across a wide range of models. In the case of single-fluid stars, this effect is described by the Hartle-Thorne equation~\cite{1967ApJ...150.1005H, 1968ApJ...153..807H}, where the frame-dragging function satisfies a second-order differential equation governed by the background TOV solution. The total moment of inertia emerges naturally from the asymptotic behavior of this function and has been widely used to study rotational properties across different nuclear equations of state.
Although rotational properties of multi-fluid systems have been addressed in recent studies--particularly in the context of rapidly rotating stars, superfluid dynamics, or related dark-sector scenarios~\cite{Konstantinou_2024, qcl7-m5kf, NAndersson_2001, refId0, PhysRevD.107.044034, Herdeiro_2024, cipriani2025differentiallyrotatingneutronstars}--a systematic analysis of the slow-rotation frame-dragging structure and its implications for effective and intrinsic moments of inertia, together with their impact on rotational–tidal universal relations within a fully gravitationally coupled two-fluid formalism remains comparatively less developed.
%Although rotational properties of multi-fluid systems have been addressed in some studies--particularly in the context of rapidly rotating stars or superfluid dynamics~\cite{Konstantinou_2024, qcl7-m5kf, NAndersson_2001, refId0, PhysRevD.107.044034}--the influence of a gravitationally coupled dark matter component on the frame-dragging profile and moment of inertia in the slow-rotation limit has not yet been examined in detail.

In this work, we investigate the slow-rotation dynamics of two-fluid neutron stars containing a gravitationally coupled dark matter component, with the aim of quantifying how dark matter microphysics influences the frame-dragging profile and total moment of inertia. We begin by deriving the rotational perturbation equation for such systems, generalizing the Hartle-Thorne formalism to the case of two distinct fluids interacting solely through gravity. The resulting frame-dragging equation is solved self-consistently together with the two-fluid TOV background, ensuring that both components contribute consistently to the rotational response. We then compute the moment of inertia across a representative set of dark matter models, including a mirror-like dark matter model and a fermionic self-interacting sector, to assess how the presence of a dark component alters the rotational structure of the star. Our analysis identifies characteristic trends in the frame-dragging profile and the resulting moment of inertia, which may serve as potential diagnostics for constraining gravitationally coupled dark matter in neutron star interiors.

The rest of this paper is organized as follows. In Sec.~\ref{sec:2}, we present the theoretical framework for slowly rotating two-fluid neutron stars. Section~\ref{sec:2a} introduces the equilibrium background equations for gravitationally coupled nuclear and dark matter fluids. The slow-rotation formalism governing the frame-dragging function is presented in Sec.~\ref{sec:2b}. In Sec.~\ref{sec:2c}, we derive the total angular momentum and moment of inertia for two-fluid configurations and discuss their physical interpretation, including observationally relevant definitions. Section~\ref{sec:2d} reformulates the rotational dynamics in terms of a basis decomposition and introduces the eigen-moments of inertia that characterize the intrinsic rotational response of the coupled system. In Sec.~\ref{sec:2e}, we summarize the nuclear matter equations of state used in this work and introduce the dark matter model, including the underlying equation of state. In Sec.~\ref{sec:3}, we present our numerical results for slowly rotating two-fluid neutron stars. We first examine the frame-dragging profiles, effective moments of inertia, and their dependence on the underlying dark matter microphysics (Sec.~\ref{sec:3a}), then analyze the intrinsic rotational modes obtained from the inertia-matrix eigenvalue decomposition (Sec.~\ref{sec:3b}). We finally examine the validity and robustness of universal relations involving the moment of inertia and tidal deformability in two-fluid configurations in the presence of a dark sector (Sec.~\ref{sec:3c}). Finally, we summarize our findings and discuss potential implications in Sec.~\ref{sec:4}.

%%%%%%%%%%%%%%%%%%%%%%%%%%%%%
\section{Theoretical Framework}
\label{sec:2}
%%%%%%%%%%%%%%%%%%%%%%%%%%%%%
%%%%%%%%%%%%%%%%%%%%%%%%%%%%%
\subsection{Equilibrium Structure of Static Stars}
%(Two-fluid TOV equations)}
\label{sec:2a}
%%%%%%%%%%%%%%%%%%%%%%%%%%%%%
We consider a two-fluid neutron star model in the framework of general relativity, where the two fluids--denoted as fluid X and fluid Y--interact solely through gravity, i.e., there is no direct non-gravitational interaction or exchange of momentum between them. Each fluid may represent a distinct physical component, such as baryonic matter and a dark matter sector. To describe the spacetime geometry, we adopt the standard line element for a static, spherically symmetric metric,
%%%%%%%%%%%%%%%
\begin{equation}
ds^{2} = -e^{2\Phi(r)} dt^{2} + e^{2\Lambda(r)} dr^{2} + r^{2} d\theta^{2} + r^{2} \sin^{2}\theta \ d\phi^{2},
\label{eq:eq1}
\end{equation}
%%%%%%%%%%%%%%%
where $\Phi(r)$ and $\Lambda(r)$ are functions of the radial coordinate $r$, representing the gravitational redshift potential and spatial curvature, respectively. The mass function $m(r)$ is introduced via $e^{2\Lambda} = \left(1-2m(r)/r\right)^{-1}$, ensuring consistency with the Schwarzschild solution in the exterior vacuum. The energy-momentum tensors of the two fluids are conserved independently, satisfying $\nabla_{\mu} T^{\mu\nu}_{X} = 0$ and $\nabla_{\mu} T^{\mu\nu}_{Y} = 0$, where the covariant derivative is defined by the metric in Eq.~(\ref{eq:eq1}). The energy-momentum tensor for each fluid is modeled as that of a perfect fluid:
%%%%%%%%%%%%%%%
\begin{equation}
T^{\mu\nu}_{i} = \left({\cal E}_{i} + P_{i}\right) u^{\mu}_{i} u^{\nu}_{i} + P_{i}\ g^{\mu\nu}, \label{eq:eq2}
\end{equation}
%%%%%%%%%%%%%%%
where $i = X, Y$ and ${\cal E}_{i}$, $P_{i}$ and $u_{i}^\mu$ represent the energy density, pressure, and four-velocity of each fluid, respectively. In the static configuration, both fluids are at rest with respect to the coordinate frame, and their four-velocities reduce to $u^{\mu}_{i} = e^{-\Phi}\ \delta^{\mu}_{t}$ with $i = X, Y$, where $\delta^{\mu}_{t}$ denotes the Kronecker delta, equal to unity for $\mu = t$ and zero otherwise. This ensures that the four-velocity has only a non-vanishing temporal component in the static background.

The condition $\nabla_{\mu} T^{\mu\nu}_{i} = 0$ provides two sets of equations for each fluid: the energy conservation and force balance (Euler) equations. The Euler equation is obtained by projecting the conservation law orthogonally to the fluid four-velocity. This yields the relativistic Euler equation for each fluid:
%%%%%%%%%%%%%%%
\begin{equation}
\frac{dP_{i}}{dr} = -\left({\cal E}_{i} + P_{i}\right) \frac{d\Phi}{dr},
\label{eq:euler}
\end{equation}
%%%%%%%%%%%%%%%
which describes hydrostatic equilibrium under the gravitational field. Since there is no direct coupling between the two fluids apart from gravity, this equation is satisfied independently for each fluid. The gravitational field is determined by Einstein's equations, $G_{\mu\nu} = 8\pi\ T_{\mu\nu}$, where the total energy-momentum tensor is the sum of the contributions from the two fluids: $T^{\mu\nu} = T^{\mu\nu}_{X} + T^{\mu\nu}_{Y}$. From the $G_{tt}$ component, we obtain the mass continuity equation,
%%%%%%%%%%%%%%%
\begin{equation}
\frac{dm}{dr} = 4\pi r^{2} \sum_{i=X, Y} {\cal E}_{i} \ ,
\label{eq:mass}
\end{equation}
%%%%%%%%%%%%%%%
which describes how the total enclosed (gravitational) mass accumulates as a function of radius. The $G_{rr}$ component yields the gradient of the gravitational potential $\Phi(r)$:
%%%%%%%%%%%%%%%
\begin{equation}
\frac{d\Phi}{dr} = \frac{1}{r (r - 2m)} \left(m + 4\pi r^{3} \sum_{i=X, Y} P_{i}\right).
\label{eq:phi}
\end{equation}
%%%%%%%%%%%%%%%
Equations \eqref{eq:euler}-\eqref{eq:phi} fully specify the background structure of a two-fluid neutron star in the static limit. The system can be integrated numerically once an equation of state is provided for each fluid.

%%%%%%%%%%%%%%%%%%%%%%%%%%%%%
\subsection{Slow-Rotation Approximation and Frame Dragging}
\label{sec:2b}
%%%%%%%%%%%%%%%%%%%%%%%%%%%%%
We now extend this model to include rotation. In the slow-rotation approximation, introduced by Hartle \cite{1967ApJ...150.1005H}, rotation is assumed to be sufficiently slow that the star remains nearly spherical, with rotational effects entering as first-order perturbations to the static background model. In this regime, the angular dependence of the metric functions can be neglected as a consequence of the first-order perturbative treatment and the regularity and boundary conditions imposed on the rotational perturbations, and the star is taken to rotate rigidly with a uniform angular velocity as measured by a distant observer. Rotation induces an off-diagonal component $g_{t\phi}$ in the metric, corresponding to the dragging of inertial frames. This effect is encoded in the frame-dragging function $\omega(r)$, defined as the local angular velocity of inertial frames relative to infinity. The perturbed metric, accurate to first order in rotation, is given by:
%%%%%%%%%%%%%%%
\begin{align}
ds^{2} =& -e^{2\Phi(r)} dt^{2} + e^{2\Lambda(r)} dr^{2} \nonumber \\
&+ r^{2} \left[d\theta^{2} + \sin^{2}\theta \left(d\phi - \omega(r) dt\right)^{2}\right],
\label{eq:rot_metric}
\end{align}
%%%%%%%%%%%%%%%
where the functions $\Phi(r)$ and $\Lambda(r)$ remain unchanged from the non-rotating (TOV) solution, and $\omega(r)$ encodes the rotational frame-dragging effect. The function $\omega(r)$ vanishes at spatial infinity, ensuring consistency with asymptotic flatness. In this approximation, the four-velocity of each fluid takes the form $u_{i}^{\mu} = e^{-\Phi} \left(1, 0, 0, \Omega_{i}\right)$, where $\Omega_{i}$ is the global angular velocity of fluid `$i$'. In the two-fluid description, the fluids are permitted to rotate rigidly with independent constant angular velocities $\Omega_{X}$ and $\Omega_{Y}$ for the inner and outer fluids, respectively. These angular velocities are treated as free parameters at this stage. Throughout this work, we label the inner fluid as $ X$, defined as the component confined within the inner boundary, and the outer fluid as $Y$, which extends from the center to the stellar surface, based purely on their radial extent, irrespective of whether the nuclear or dark matter component occupies the inner region.

Under the slow-rotation approximation, the frame-dragging function $\omega(r)$ is governed by the $t\phi$-component of Einstein’s equations, which connects the rotational perturbation to the energy and pressure distributions of the star. Evaluating the Einstein tensor for the metric in Eq.~\eqref{eq:rot_metric}, and retaining only terms linear in $\omega$ (consistent with the slow-rotation limit), the $G_{t\phi}$ component becomes:
%%%%%%%%%%%%%%%
\begin{align}
    G_{t\phi} =& \left[0.5r^{2}\frac{d^2\omega}{dr^{2}} + r\left\{2-0.5r\left(\frac{d\Phi}{dr} + \frac{d\Lambda}{dr}\right)\right\}\frac{d\omega}{dr} \right. \nonumber \\
    & \left. -r^{2}\omega\left\{\frac{d^{2}\Phi}{dr^{2}}+\left(\frac{d\Phi}{dr}\right)^{2}-\frac{d\Phi}{dr}\frac{d\Lambda}{dr}\right\} \right. \nonumber \\
    & \left. + r\omega\left(\frac{d\Lambda}{dr} - \frac{d\Phi}{dr}\right) \right] e^{-2\Lambda}\sin^{2}\theta
    \label{eq:eqG}
\end{align}
%%%%%%%%%%%%%%%
%\begin{align}
%G_{t\phi} =& \frac{1}{2}r^{2}e^{-2\Lambda}\sin^{2}\theta\ \frac{d^2\omega}{dr^{2}} + 2re^{-2\Lambda}\sin^{2}\theta\ \frac{d\omega}{dr} \nonumber \\
%& -\frac{1}{2}r^{2}e^{-2\Lambda}\sin^{2}\theta\left(\frac{d\Phi}{dr} + \frac{d\Lambda}{dr}\right)\frac{d\omega}{dr} \nonumber \\
%& -8\pi Pr^{2} \omega \sin^{2}\theta,
%\label{eq:eqG}
%\end{align}
%%%%%%%%%%%%%%%
Substituting $G_{t\phi}$ and $T_{t\phi}$ into Einstein’s equation ($G_{\mu\nu} = 8\pi\, T_{\mu\nu}$), and simplifying using the background (non-rotating) field equations for $\Phi(r)$ and $\Lambda(r)$, we obtain the master equation governing the radial profile of the frame-dragging function $\omega(r)$ in a slowly rotating, two-fluid neutron star, with explicit dependence on the energy densities, pressures, and angular velocities of both fluids:
%%%%%%%%%%%%%%%
\begin{widetext}
\begin{equation}
r^{2} \frac{d^{2}\omega}{dr^{2}} + \left[ 4r - r^{2} \left( \frac{d\Phi}{dr} + \frac{d\Lambda}{dr} \right) \right] \frac{d\omega}{dr} + 16\pi e^{2\Lambda} r^{2} \Big[ \left( {\cal E}_{X} + P_{X} \right) \left( \Omega_{X} - \omega \right) + \left( {\cal E}_{Y} + P_{Y} \right) \left( \Omega_{Y} - \omega \right) \Big] = 0\ .
\label{eq:masterEq}
\end{equation}
\end{widetext}
%%%%%%%%%%%%%%%
%Equation \eqref{eq:masterEq} describes the radial behavior of the frame-dragging angular velocity $\omega(r)$ in a slowly rotating, two-fluid neutron star, with explicit dependence on the energy densities, pressures, and rotation rates of both fluids. 
In the case of a single-fluid star, Eq.~\eqref{eq:masterEq} reduces to the standard Hartle-Thorne frame-dragging equation~\cite{1968ApJ...153..807H}.

To solve the frame-dragging equation, appropriate boundary conditions must be imposed. At the center of the star, regularity requires the derivative of the frame-dragging function to vanish, i.e., $\left(d\omega/dr\right)_{r=0} = 0$, ensuring smoothness and finiteness of the solution. At the stellar surface--defined by the outer boundary of fluid $Y$ in the two-fluid model---the interior solution must be smoothly matched to the exterior vacuum solution.  In the exterior region, where both fluids vanish 
(${\cal E}_{X} = {\cal E}_{Y} = P_{X} = P_{Y} = 0$), the master equation [Eq. \eqref{eq:masterEq}] reduces to a homogeneous form whose general solution behaves as $\omega(r) \propto r^{-3}$. This asymptotic behavior ensures that the frame-dragging effect decays at large distances, consistent with asymptotic flatness. Matching the interior and exterior solutions across the stellar surface leads to the continuity condition on the metric~\cite{PhysRevD.86.124036}
%%%%%%%%%%%%%%%
\begin{equation}
r \frac{d\omega}{dr} + 3\ \omega = 0,
\end{equation}
%%%%%%%%%%%%%%%
to be satisfied at $r=R_{Y}$, where $R_{Y}$ denotes the radius of outer fluid $Y$. 

The numerical solution of the master equation is obtained using a standard shooting method. The equation is integrated outward from the center of the star, beginning with a trial central value $\omega(0) = \omega_{c}$. This central value is iteratively adjusted until the solution satisfies the surface boundary condition at $R_{Y}$, ensuring a smooth match to the exterior vacuum solution. In this work, we consider both co-rotating configurations, where $\Omega_{X} = \Omega_{Y}$, and non-co-rotating cases ($\Omega_{X} \neq \Omega_{Y}$) in which the two fluids rotate with distinct angular velocities. This enables us to systematically assess how relative rotation between the two fluids affects the frame-dragging profile and the resulting spacetime structure. Throughout our analysis, fluid $Y$ defines the overall stellar surface, and the outer boundary condition is enforced at $R_{Y}$.

An important structural feature of the frame-dragging equation [Eq.~\eqref{eq:masterEq}] is its linearity. Specifically, the master equation is linear in the frame-dragging function $\omega(r)$, with the rotation rates $\Omega_{X}$ and $\Omega_{Y}$ appearing only as inhomogeneous source terms. This structure allows the equation to be recast in a more transparent and modular form:
%%%%%%%%%%%%%%%
\begin{equation}
    \frac{d^{2}\omega}{dr^{2}} + A(r) \frac{d\omega}{dr} + B(r)\, \omega = S_{X}(r)\, \Omega_{X} + S_{Y}(r)\, \Omega_{Y}
    \label{scheMasterEq}
\end{equation}
%%%%%%%%%%%%%%%
with coefficients:
%%%%%%%%%%%%%%%
\begin{eqnarray}
    A(r) &=&  \frac{4}{r} - \frac{d\Phi}{dr} - \frac{d\Lambda}{dr}, \nonumber \\
    B(r) &=&  - 16\pi e^{2\Lambda} \left( {\cal E}_{X} + P_{X} + {\cal E}_{Y} + P_{Y} \right), \nonumber \\
    S_{X}(r) &=& - 16\pi e^{2\Lambda} \left( {\cal E}_{X} + P_{X} \right), \nonumber \\
    S_{Y}(r) &=& - 16\pi e^{2\Lambda} \left( {\cal E}_{Y} + P_{Y} \right), \nonumber 
\end{eqnarray}
%%%%%%%%%%%%%%%
which depend solely on the background structure of the star--i.e., the static metric and fluid profiles in the non-rotating configuration.

Since the coefficients $A$, $B$, $S_X$, and $S_Y$ are independent of the rotation rates, the solution $\omega(r)$ for a fixed background configuration can be expressed as a linear superposition of basis responses to the rotation of each fluid:
%%%%%%%%%%%%%%%
\begin{equation}
\omega(r) = a_X(r)\, \Omega_X + a_Y(r)\, \Omega_Y,
\label{eq:omega_solution}
\end{equation}
%%%%%%%%%%%%%%%
where $a_{X}(r)$ and $a_{Y}(r)$ are basis functions obtained by solving Eq.~\eqref{scheMasterEq} for the two independent choices $(\Omega_{X}, \Omega_{Y}) = (1, 0)$ and $(0, 1)$, with the same boundary conditions imposed at the center and at $r = R_Y$. These basis functions are computed on the full two-fluid background and therefore incorporate the influence of both fluids, even when only one is assigned a unit rotation rate. Moreover, the outer boundary condition at the stellar surface, $r\, (d\omega/dr) + 3\omega = 0$, is also linear in $\omega$ and thus preserves the validity of the decomposition [Eq. \eqref{eq:omega_solution}] all the way up to and including the stellar surface.

This formulation significantly simplifies both the numerical implementation and the physical interpretation of the problem. Once the background stellar structure is fixed, the full frame-dragging profile for any choice of $(\Omega_X,\Omega_Y)$ can be constructed directly from the precomputed basis functions $a_X$ and $a_Y$. In what follows, we leverage this decomposition to systematically analyze the angular momentum and moment of inertia of two-fluid stars in both co-rotating and non-co-rotating scenarios.

We emphasize that in the present work the two fluids are modeled as independent perfect fluids coupled exclusively through gravity, with no entrainment or additional microphysical interactions. In particular, the stress–energy tensor is a simple sum of the individual perfect-fluid contributions, and each component satisfies its own conservation law. The slow-rotation scheme adopted here therefore corresponds to a direct first-order Hartle–Thorne expansion of this additive system. The additional structural features that arise in relativistic superfluid multi-fluid formalisms with entrainment (see, e.g., Ref.~\cite{PhysRevD.107.044034}) are absent in the present model. Consequently, the standard first-order slow-rotation treatment remains self-consistent for the class of two-fluid configurations considered here.

%%%%%%%%%%%%%%%%%%%%%%%%%%%%%
\subsection{Angular Momentum and Moment of Inertia}
\label{sec:2c}
%%%%%%%%%%%%%%%%%%%%%%%%%%%%%
Once the frame-dragging function $\omega(r)$ is known, the total angular momentum carried by the rotating two-fluid star can be computed. In general relativity, the angular momentum is defined through a volume integral of the mixed stress-energy component $T^{t}_{\ \phi}$ and can be expressed as~\cite{1971ApJ...167..359B, 1973Ap&SS..24..385H, PhysRevD.50.3836},
%%%%%%%%%%%%%%%
\begin{equation}
J = \int T^{t}_{\ \phi}\ \sqrt{-g}\ dr\ d\theta\ d\phi,
\end{equation}
%%%%%%%%%%%%%%%
where $\sqrt{-g}$ is the metric determinant associated with the slowly rotating metric in Eq.~\eqref{eq:rot_metric}. 

For a perfect fluid, the relevant component of the stress-energy tensor reduces to $T^{t}_{\ \phi} = ({\cal E} + P) u^{t} u_{\phi}$, which is linear in the fluid angular velocity in the slow-rotation limit. Since the two fluids are allowed to rotate rigidly with independent angular velocities $\Omega_{X}$ and $\Omega_{Y}$,  the mixed stress-energy component entering the angular momentum integral is given by the sum of the individual fluid contributions. Consequently, the total angular momentum receives separate contributions from each component, while remaining coupled at the level of rotational perturbations through the common frame-dragging function $\omega(r)$. Substituting the four-velocities into $T^{t}_{\ \phi}$, and carrying out the angular integrations, the total angular momentum of the two-fluid star can be written as
%%%%%%%%%%%%%%%
\begin{align}
J_{T} =& \, \frac{8\pi}{3} \int_{0}^{R_{Y}}\ \Big[ \left( {\cal E}_X + P_X \right) \left( \Omega_X - \omega \right) \nonumber \\
&+ \left( {\cal E}_Y + P_Y \right) \left( \Omega_Y - \omega \right) \Big]\ r^{4}\ e^{\Lambda - \Phi}\ dr\ .
\label{eq:totalAngMom}
\end{align}
%%%%%%%%%%%%%%%
The radial integration in Eq.~\eqref{eq:totalAngMom} extends from the stellar center up to the outer radius $R_{Y}$, which defines the surface of the two-fluid configuration. While the explicit contribution of each fluid to the integrand is confined to the region where its energy density and pressure are nonzero, the frame-dragging profile $\omega(r)$ encodes the rotational response of the spacetime arising from the gravitational coupling between the two fluids, thereby ensuring that the total angular momentum consistently accounts for both the direct contributions of each component and their coupling at the linear level of rotational perturbations.

Now, substituting the basis decomposition of the frame-dragging function, $\omega = a_X\Omega_X + a_Y\Omega_Y$, into the expression for the total angular momentum $J_{T}$ in Eq.~\eqref{eq:totalAngMom}, the integrand reorganizes into terms proportional to $\Omega_{X}$ and $\Omega_{Y}$. This makes the linear dependence of $J_{T}$ on the rotation-rate pair $(\Omega_{X}, \Omega_{Y})$ explicit, so the resulting expression can be written as
%%%%%%%%%%%%%%%
\begin{widetext}
    \begin{align}
        J_{T} =&\, \underbrace{\frac{8\pi}{3} \int_0^{R_Y} \Big[\left({\cal E}_X + P_X\right) \left(1 - a_X\right) + \left({\cal E}_Y + P_Y\right)\,(-a_{X}) \Big]\,  r^{4}e^{\Lambda-\Phi} dr}_{\displaystyle = I_X^{\rm eff}}\ {\Omega_{X}} \nonumber \\
        &+ \underbrace{\frac{8\pi}{3} \int_0^{R_Y} \Big[\left({\cal E}_X + P_X\right) \left(- a_Y\right) + \left({\cal E}_Y + P_Y\right)\,(1-a_{Y}) \Big]\,  r^{4}e^{\Lambda-\Phi} dr}_{\displaystyle = I_Y^{\rm eff}} \ {\Omega_{Y}}\ .
        \label{eq:JT_linearOmega}
    \end{align}
\end{widetext}
%%%%%%%%%%%%%%%
The coefficients $I_{X}^{\rm{eff}}$ and $I_{Y}^{\rm{eff}}$ defined in Eq.~\eqref{eq:JT_linearOmega} represent the effective moments of inertia associated with the rigid rotation rates $\Omega_{X}$ and $\Omega_{Y}$, respectively. They quantify the angular momentum generated per unit $\Omega_{X}\ {\rm{or}}\ \Omega_{Y}$ for a fixed two-fluid background configuration, once the rotational response of the spacetime is fully taken into account. This response is encoded in the frame-dragging basis functions $a_{X}(r)$ and $a_{Y}(r)$, which are obtained by solving the frame-dragging equation on the coupled two-fluid background. As a result, both $I_{X}^{{\rm{eff}}}$ and $I_{Y}^{{\rm{eff}}}$ intrinsically incorporate the gravitationally mediated rotational coupling between the two components.

To make this structure explicit, it is convenient to separate the effective inertias into contributions associated with each fluid and each basis response. We therefore define the self- and cross-coupled inertia coefficients
%%%%%%%%%%%%%%%
\begin{subequations}
\begin{align}
I_{XX} & \equiv  \frac{8\pi}{3} \int_{0}^{R_{Y}} \left({\cal E}_{X} + P_{X}\right) \left(1 - a_{X}\right)\, r^{4}e^{\Lambda-\Phi} dr, \label{eq:15a} \\
I_{YX} & \equiv  \frac{8\pi}{3} \int_{0}^{R_{Y}} \left({\cal E}_{Y} + P_{Y}\right) \left(- a_{X}\right)\, r^{4}e^{\Lambda-\Phi} dr, \label{eq:15b} \\
I_{XY} & \equiv  \frac{8\pi}{3} \int_{0}^{R_{Y}} \left({\cal E}_{X} + P_{X}\right) \left(- a_{Y}\right)\, r^{4}e^{\Lambda-\Phi} dr, \label{eq:15c} \\
I_{YY} & \equiv  \frac{8\pi}{3} \int_{0}^{R_{Y}} \left({\cal E}_{Y} + P_{Y}\right) \left(1 - a_{Y}\right)\, r^{4}e^{\Lambda-\Phi} dr. \label{eq:15d} 
\end{align}
\label{eq:inertia_coeff}
\end{subequations}
%%%%%%%%%%%%%%%
The coefficients $I_{ij}$ defined above admit a transparent physical interpretation. The first index $i$ identifies the fluid whose stress-energy contributes to the angular momentum density, while the second index $j$ specifies the rotational basis function $a_{j}(r)$ through which the frame-dragging response enters. Accordingly, the diagonal terms $I_{XX}$ and $I_{YY}$ are associated with the angular momentum carried by fluids $X$ and $Y$, respectively, in response to their own rotation, with the factors $(1-a_{X})$ and $(1-a_{Y})$ encoding the relativistic reduction due to frame dragging. The off-diagonal terms $I_{XY}$ and $I_{YX}$, on the other hand, quantify the angular momentum induced in one fluid by the frame-dragging field sourced by the rotation of the other fluid. These cross terms arise because both components source and respond to the same spacetime geometry, so that the rotation of either fluid generates a global frame-dragging response that necessarily affects the angular momentum carried by the other component. In particular, even when the outer fluid has zero rotation rate ($\Omega_Y = 0$), it can still carry angular momentum through the cross term $I_{YX}$, reflecting the frame-dragging response induced by the rotation of fluid $X$.

By construction, the effective inertias appearing in Eq.~\eqref{eq:JT_linearOmega} decompose as 
%%%%%%%%%%%%%%%
\begin{equation}
    I_{X}^{{\rm{eff}}} = I_{XX} + I_{YX}\, , \quad \quad \quad I_{Y}^{{\rm{eff}}} = I_{YY} + I_{XY}\, ,
    \label{eq:Ieff_decomp}
\end{equation}
%%%%%%%%%%%%%%%
and the total angular momentum can be expressed compactly as
%%%%%%%%%%%%%%%
\begin{equation}
    J_{T} = \sum_{i, j\, \in\{X, Y\}} I_{ij}\, \Omega_{j}.
    \label{eq:JT_Iij}
\end{equation}
%%%%%%%%%%%%%%%
Although $\Omega_{X}$ and $\Omega_{Y}$ enter Eq.~\eqref{eq:JT_Iij} as independent rotation rates, the coefficients $I_{ij}$ are not free parameters: for any fixed two-fluid background configuration, they are determined uniquely by the equilibrium profiles and by the basis functions $a_{X}(r)$ and $a_{Y}(r)$, which encode the frame-dragging response of the coupled spacetime.

It is worth emphasizing that $I_{X}^{\rm eff}$ and $I_{Y}^{\rm eff}$ should not be interpreted as the moments of inertia of the individual fluids in isolation. Each effective inertia generally contains contributions from both components through the common frame-dragging response. However, since $I_{X}^{\rm eff}$ and $I_{Y}^{\rm eff}$ depend only on the static stellar structure and the associated frame-dragging basis functions, their sum,
%%%%%%%%%%%%%%%
\begin{equation}
I_{T}^{\rm eff} \equiv I_{X}^{\rm eff} + I_{Y}^{\rm eff} \, ,
\label{eq:ITeff}
\end{equation}
%%%%%%%%%%%%%%%
provides a natural and physically meaningful definition of the total moment of inertia of the two-fluid configuration within the slow-rotation approximation. This quantity characterizes the global rotational response of the star as a whole and is uniquely determined by the equilibrium background, independent of the particular values of $\Omega_{X}$ and $\Omega_{Y}$. As such, $I_{T}^{\rm{eff}}$ represents the appropriate generalization of the moment of inertia for two-fluid relativistic stars and therefore constitutes the relevant quantity for studies of universal relations involving rotational properties.

In practice, astrophysical observations probe the rotational state of a neutron star through electromagnetic emission from its baryonic surface, yielding a single, externally measured spin frequency associated with the rotation of nuclear matter. Within the present two-fluid framework, this observed spin frequency is associated with the rotation of the nuclear (baryonic) component, since electromagnetic emission originates from the visible stellar surface. However, even when the two fluids are coupled only through gravity, their rotational dynamics are not entirely independent, as both components contribute to the global spacetime structure and hence affect the overall rotational dynamics through the common gravitational field. Consequently, while the measured frequency reflects the rotation of nuclear matter, it remains influenced indirectly by the presence and distribution of the dark component. The internal rotational state of the individual fluids cannot be resolved observationally through the spin frequency alone, and the two fluids may therefore rotate at different angular velocities, $\Omega_X \neq \Omega_Y$.

It is therefore natural to define an observationally relevant moment of inertia by normalizing the total angular momentum by the rotation rate of the nuclear (visible) component,
%%%%%%%%%%%%%%%
\begin{equation}
    I_{\rm{obs}} \equiv \frac{J_{T}}{\Omega_{_{\rm{NM}}}},
    \label{eq:eq19}
\end{equation}
%%%%%%%%%%%%%%%
where $\Omega_{\rm{NM}}$ denotes the angular velocity of nuclear matter. In configurations where the dark matter forms a core and nuclear matter extends to the stellar surface, one has $\Omega_{\rm{NM}} = \Omega_{Y}$, and the definition reduces to
%%%%%%%%%%%%%%%
\begin{equation}
    I_{\rm{obs}} \equiv \frac{J_{T}}{\Omega_{Y}} = I_{Y}^{{\rm{eff}}} + \left(\frac{\Omega_{X}}{\Omega_{Y}}\right)I_{X}^{{\rm{eff}}}.
    \label{eq:eq20}
\end{equation}
%%%%%%%%%%%%%%%
In contrast, for dark-matter halo configurations where nuclear matter is confined to an inner region and dark matter occupies the outer layers, the observed spin still corresponds to the rotation of the nuclear component, $\Omega_{\rm{NM}} = \Omega_{X}$, so that
%%%%%%%%%%%%%%%
\begin{equation}
    I_{\rm{obs}} \equiv \frac{J_{T}}{\Omega_{X}} = \left(\frac{\Omega_{Y}}{\Omega_{X}}\right) I_{Y}^{{\rm{eff}}} + I_{X}^{{\rm{eff}}}.
    \label{eq:eq21}
\end{equation}
%%%%%%%%%%%%%%%

In both cases, $I_{\rm{obs}}$ represents the effective moment of inertia that would be inferred from rotational observables when observers measure a single global spin frequency and interpret it as arising from a rigidly rotating star, without access to the internal multi-fluid structure, while the definition itself fully encodes the contributions of both fluids through the total angular momentum $J_{T}$, ensuring that $I_{\rm{obs}}$ captures the cumulative rotational response of the system as ``seen" by a distant observer. In the co-rotating limit $(\Omega_{X} = \Omega_{Y})$, this quantity reduces smoothly to the effective total moment of inertia, $I_{\rm{obs}} = I_{T}^{{\rm{eff}}}$, and depends solely on the equilibrium stellar structure. As such, $I_{\rm{obs}}$ provides the appropriate bridge between the two-fluid formalism developed here and observationally inferred rotational properties, while clearly distinguishing between core- and halo-type configurations.
%%%%%%%%%%%%%%%%%%%%%%%%%%%%%
\subsection{Basis decomposition and Eigen-moments of Inertia}
\label{sec:2d}
%%%%%%%%%%%%%%%%%%%%%%%%%%%%%
The formulation developed in Sec.~\ref{sec:2c} shows that, within the slow-rotation approximation, the total angular momentum of a gravitationally coupled two-fluid relativistic star depends linearly on the pair of rigid rotation rates $(\Omega_{X},\, \Omega_{Y})$. The linear relation between the total angular momentum and the rotation rates derived in Eq.~\eqref{eq:JT_Iij} can be written in a compact matrix form by collecting the inertia coefficients into a $2\times2$ inertia matrix,
%%%%%%%%%%%%%%%
\begin{equation}
\label{eq:inertia_matrix}
\mathbf{I} \equiv
\begin{pmatrix}
I_{XX} & I_{XY} \\
I_{YX} & I_{YY}
\end{pmatrix}.
\end{equation}
%%%%%%%%%%%%%%%
For a given two-fluid equilibrium configuration, this matrix is completely determined by the background structure and the associated frame-dragging basis functions. The total angular momentum can then be obtained by first applying the inertia matrix to the rotation-rate pair and then summing the resulting angular-momentum contributions,
%The total angular momentum can then be expressed as a bilinear form involving the inertia matrix and the rotation-rate pair,
%%%%%%%%%%%%%%%
\begin{equation}
\label{eq:J_matrix_form}
J_T = \sum_{i,j\in{X,Y}} I_{ij}\,\Omega_j
=
\begin{pmatrix}
1 & 1
\end{pmatrix}
\, \mathbf{I} \,
\begin{pmatrix}
\Omega_X \\ \Omega_Y
\end{pmatrix}.
\end{equation}
%%%%%%%%%%%%%%%
In this form, the rotational response of the two-fluid star is naturally interpreted as the action of a linear inertia operator on the two-dimensional space spanned by the independent rotation rates $(\Omega_{X},\, \Omega_{Y})$. Since the presence of gravitational coupling implies that this operator is not generally diagonal in the fluid-rotation basis, the individual angular velocities $\Omega_{X}$ and $\Omega_{Y}$ do not, in general, represent the intrinsic rotational degrees of freedom of the system. This observation motivates the introduction of a rotated basis in which the inertia matrix is diagonal, allowing the coupled rotational dynamics to be described in terms of independent eigenmodes characterized by well-defined ``eigen-moments of inertia''.

The intrinsic rotational degrees of freedom of the coupled system are most transparently exposed by diagonalizing the inertia operator $\mathbf{I}$. Since $\mathbf{I}$ is a real $2\times2$ matrix determined entirely by the equilibrium background, it admits a spectral decomposition in terms of its eigenvalues and eigenvectors. Introducing a linear transformation in rotation-rate space,
%%%%%%%%%%%%%%%
\begin{equation}
    \begin{pmatrix}
        \Omega_{+} \\ \Omega_{-}
    \end{pmatrix}
    = \mathbf{P^{-1}} \begin{pmatrix}
        \Omega_{X} \\ \Omega_{Y}
    \end{pmatrix}
    ,
\end{equation}
%%%%%%%%%%%%%%%
where the columns of $\mathbf{P}$ are the eigenvectors of the inertia matrix $\mathbf{I}$, the inertia operator can be brought into diagonal form,
%%%%%%%%%%%%%%%
\begin{equation}
    \mathbf{P^{-1}\, I\, P} = \begin{pmatrix}
        I_{+} & 0 \\
        0 & I_{-} 
    \end{pmatrix}.
\end{equation}
%%%%%%%%%%%%%%%
The eigenvectors defining $\mathbf{P}$ are normalized such that the total angular momentum retains its additive and linear functional form under the change of basis. Such a normalization is always possible since the eigenvectors of a finite-dimensional linear operator are defined up to an arbitrary nonzero scaling, and it does not affect the diagonalization of the inertia matrix. Explicitly, we impose the condition
%%%%%%%%%%%%%%%
\begin{equation}
    \begin{pmatrix}1 & 1\end{pmatrix}\mathbf{P}
    =
    \begin{pmatrix}1 & 1\end{pmatrix},
\end{equation}
%%%%%%%%%%%%%%%
which ensures that the expression for the total angular momentum as a linear combination of rotation variables is invariant under the basis transformation. This normalization uniquely fixes the otherwise arbitrary scaling of the eigenvectors and guarantees that the total angular momentum is preserved as a linear functional of the rotation variables in both the original fluid basis and the eigen-rotation basis. In this rotated basis, the total angular momentum decomposes into independent contributions proportional to the new rotation variables $\Omega_{+}$ and $\Omega_{-}$, each weighted by a corresponding eigenvalue $I_{+}$ or $I_{-}$. These eigenvalues define the eigen-moments of inertia of the two-fluid star, while $\Omega_{+}$ and $\Omega_{-}$ represent the associated collective rotation patterns. Unlike the original fluid rotation rates, which are coupled through frame dragging, the eigen-rotation variables isolate the fundamental rotational modes of the gravitationally coupled system and provide a natural basis for characterizing its rotational dynamics.

The eigenvalues $I_{+}$ and $I_{-}$ are obtained by solving the characteristic equation of the inertia matrix $\mathbf{I}$ with respect to $\lambda$, i.e., $\det\,(\mathbf{I}- \lambda\, {\mathcal I}) = 0$ (where ${\mathcal I}$ is identity matrix), which yields two real solutions,
%%%%%%%%%%%%%%%
\begin{equation}
    I_{\pm} = \frac{1}{2} \left[I_{XX} + I_{YY} \pm \sqrt{\left(I_{XX}-I_{YY}\right)^{2} + 4\, I_{XY} I_{YX}}\ \right].
\end{equation}
%%%%%%%%%%%%%%%
These quantities depend only on the equilibrium stellar structure and the associated frame-dragging basis functions, and are therefore fixed for a given two-fluid background configuration. By construction, $I_{+}$ and $I_{-}$ provide a complete characterization of the rotational response of the system within the slow-rotation approximation, in the sense that any rigid rotation state can be expressed as a superposition of the two corresponding eigen-rotation modes. Importantly, this decomposition is independent of the particular choice of fluid rotation rates $(\Omega_{X},\, \Omega_{Y})$ and reflects an intrinsic property of the gravitationally coupled two-fluid star.

The eigen-rotation variables $\Omega_{+}$ and $\Omega_{-}$ introduced above are linear combinations of the original fluid rotation rates $\Omega_{X}$ and $\Omega_{Y}$. Explicitly, they may be written as
%%%%%%%%%%%%%%%
\begin{equation}
    \Omega_{+} = p_{1} \Omega_{X} + p_{2}\Omega_{Y}\, , \quad \quad \quad \Omega_{-} = q_{1} \Omega_{X} + q_{2}\Omega_{Y}\, ,
\end{equation}
%%%%%%%%%%%%%%%
where the coefficients $(p_{1}, p_{2})$ and $(q_{1}, q_{2})$ are determined by the normalized eigenvectors of the inertia matrix $\mathbf{I}$ corresponding to the eigenvalues $I_{+}$ and $I_{-}$, respectively. These coefficients depend only on the background stellar structure through the inertia coefficients $I_{ij}$, and are therefore fixed for a given two-fluid equilibrium configuration. The transformation from $\Omega_{X}$ and $\Omega_{Y}$ to $\Omega_{+}$ and $\Omega_{-}$ thus represents a change of basis in rotation-rate space, in which the coupled rotational dynamics of the system are decomposed into two independent collective modes.

The eigen-rotation modes defined by $\Omega_{+}$ and $\Omega_{-}$ do not correspond to the rotation of individual fluids, but instead represent collective rotational patterns of the gravitationally coupled two-fluid system. Each eigenmode describes a coherent combination of the fluid rotation rates that excites a specific frame-dragging response of the spacetime, characterized by a single eigen-moment of inertia. In this sense, the eigenmodes capture how the star as a whole responds to rotation once the gravitational coupling between the components is taken into account. The associated eigenvalues $I_{+}$ and $I_{-}$ quantify the efficiency with which these collective rotation patterns generate angular momentum, independently of how the rotation is partitioned between the two fluids at the microscopic level.

Expressed in the eigen-rotation basis, the total angular momentum acquires a particularly simple form. Substituting the inverse transformation from $(\Omega_{X},\, \Omega_{Y})$ to $(\Omega_{+},\, \Omega_{-})$ into Eq.~\eqref{eq:JT_Iij}, and using the diagonal representation of the inertia matrix, the total angular momentum can be written as
%%%%%%%%%%%%%%%
\begin{equation}
    J_{T} = I_{+}\, \Omega_{+}\, +\, I_{-}\, \Omega_{-}.
\end{equation}
%%%%%%%%%%%%%%%
In this representation, the two collective rotation modes contribute additively and independently to the angular momentum, each weighted by its corresponding eigen-moment of inertia. This diagonal form makes explicit that, within the slow-rotation approximation, the rotational response of the coupled two-fluid star decomposes into two uncoupled linear channels, fully characterized by $(I_{+},\, \Omega_{+})$ and $(I_{-},\, \Omega_{-})$.

%%%%%%%%%%%%%%%%%%%%%%%%%%%%%
\subsection{Equation of State}
\label{sec:2e}
%%%%%%%%%%%%%%%%%%%%%%%%%%%%%
To construct self-consistent models of rotating two-fluid neutron stars, a specification of the equation of state for each constituent is required. In the present framework, the star is modeled as a gravitationally coupled system composed of ordinary nuclear matter and an additional dark matter component, each described as a perfect fluid and interacting only through the spacetime geometry. The equation of state of each fluid provides the necessary closure relation between its energy density and pressure, thereby determining both the equilibrium stellar structure and the associated rotational response within the slow-rotation approximation.

For the nuclear matter component, we adopt a set of representative relativistic equations of state that span a broad range of stiffness, satisfy key empirical constraints at and around nuclear saturation density, and are commonly used in contemporary studies of neutron star structure. Specifically, we consider three nuclear matter equation of state models: QMC-RMF4~\cite{PhysRevC.106.055804}, DD2~\cite{PhysRevC.81.015803, PhysRevC.89.064321}, and QHC21-BT~\cite{Kojo_2022, TOGASHI201778}. These models differ in their microscopic treatment of nuclear interactions and the behavior of matter at supranuclear densities, thereby allowing us to assess the robustness of our results against uncertainties in the nuclear sector. In particular, the QHC21-BT equation of state model includes quark matter in a crossover manner in the higher-density region. For non-rotating configurations, these models support maximum gravitational masses in the range $M_{\rm{max}} \simeq 2.20 - 2.42\ M_{\odot}$, with corresponding radii of order $R \simeq 10.95 - 11.95$ km~\cite{jksq-q6ty}, and predict distinct radii for canonical $1.4\ M_{\odot}$ neutron stars, $R_{1.4} \simeq 12.19$ km (QMC-RMF4), $13.16$ km (DD2), and $11.84$ km (QHC21-BT), thereby accommodating the existence of heavy neutron stars while exhibiting a controlled spread in stiffness across phenomenologically relevant density regimes.

The selection of these three equations of state is intended to deliberately span distinct classes of high-density microphysics — including conventional relativistic mean-field, density-dependent relativistic mean-field, and hybrid quark-hadron crossover models — thereby providing a controlled yet diverse sampling of phenomenologically viable nuclear matter descriptions. This choice was made a priori to probe the robustness of the rotational–tidal correlations across qualitatively different microscopic realizations, rather than to optimize the behaviour of any specific relation.

As a baseline scenario for the dark matter component, we consider mirror dark matter~\cite{Foot:2004pa, doi:10.1142/S0217751X14300130}, in which the dark matter fluid is modeled using the same equation of state as ordinary nuclear matter. In this phenomenological setup, the dark component is treated as a second baryonic-like fluid whose pressure-energy density relation mirrors that of nuclear matter, while the two fluids interact only through gravity. Consequently, the nuclear and dark matter equations of state are identical in functional form and differ only through their assigned central densities. This construction provides a particularly clean reference model, as it isolates the effects of gravitational coupling and fluid stratification without introducing additional dark-sector microphysical parameters. Mirror dark matter, therefore, serves as a useful benchmark for assessing the generic impact of a second fluid on neutron star structure and rotational properties.

Beyond this baseline scenario, we also consider a more general class of self-interacting dark matter models in which the dark component is described by a fermionic fluid with repulsive vector self-interactions~\cite{Kumar2025, Nelson_2019}. In this framework, the dark matter equation of state is obtained within a relativistic mean-field treatment of a Dirac fermion of mass $m_{\chi}$ coupled to a light vector mediator of mass $m_{\rm{v}}$ with coupling strength $g_{\chi}$. The resulting expressions for the energy density and pressure depend explicitly on the dark matter particle mass and the effective strength of the vector interaction, thereby introducing controlled microphysical parameters that regulate the stiffness of the dark matter equation of state. Specifically, the energy density and pressure of dark matter take the form~\cite{Kumar2025}
%%%%%%%%%%%%%%%
\begin{align}
    {\cal E}_{\rm{DM}} &= \frac{2}{\left(2\pi\right)^{3}} \int^{k_{\chi}^{F}}_{0} \sqrt{k^{2}+m_{\chi}^{2}}\, d^{3}k \,+\,  \frac{1}{2} \left(\frac{g_{\chi}}{m_{\rm{v}}}\right)^{2} n_{\chi}^{2}, \nonumber \\
    P_{\rm{DM}} &= \frac{2}{3\left(2\pi\right)^{3}} \int^{k_{\chi}^{F}}_{0} \frac{k^{2}}{\sqrt{k^{2}+m_{\chi}^{2}}}\, d^{3}k \,+\,  \frac{1}{2} \left(\frac{g_{\chi}}{m_{\rm{v}}}\right)^{2} n_{\chi}^{2},
    \label{eq:eq2}
\end{align}
%%%%%%%%%%%%%%%
where $k_{\chi}^{\rm{F}}$ and $n_{\chi}$ denote the Fermi momentum and number density of the dark fermions, respectively. The first terms correspond to the kinetic contribution of a degenerate fermion gas, while the second terms encode the repulsive vector self-interaction. The stiffness of the dark-matter equation of state is therefore regulated by the dark matter particle mass $m_{\chi}$ and the coupling ratio $g_{\chi}/m_{\rm{v}}$, allowing for a controlled exploration of dark-sector microphysics. This vector-mediated fermionic dark matter model provides a theoretically well-defined extension beyond mirror dark matter and enables us to assess how dark-matter self-interactions influence the equilibrium structure and rotational properties of neutron stars within the two-fluid framework. Unless stated otherwise, both the mirror dark matter and vector-mediated fermionic dark matter equations of state are employed in the two-fluid calculations presented in the following sections.

The two dark-sector scenarios considered here are not intended to exhaust the full range of dark matter candidates, but rather to represent theoretically motivated and structurally distinct classes relevant for compact star modeling. Mirror dark matter corresponds to symmetric extensions of the Standard Model in which the dark sector mirrors baryonic microphysics, providing a minimal gravitationally coupled two-fluid benchmark. The vector-mediated fermionic model belongs to the broader class of self-interacting dark matter frameworks in which repulsive interactions generate an independent stiffness scale. Many other dark matter candidates relevant for compact stars — including asymmetric or composite fermionic dark matter models — can also be described effectively as barotropic fluids and would fall within the same structural categories considered here. Together, these two limits allow us to probe whether rotational and tidal properties are sensitive primarily to the mere presence of an additional gravitating fluid or to the emergence of genuinely distinct dark-sector microphysics.

%%%%%%%%%%%%%%%
%%%%%%%%%%%%%%%%%%%%%%%%%%%%%
\section{Results and Discussion}
\label{sec:3}
%%%%%%%%%%%%%%%
In this section, we investigate the rotational response of two-fluid neutron stars within the slow-rotation framework developed in Sec.~\ref{sec:2}. We focus on how the frame-dragging profile $\omega(r)$ and the associated moments of inertia depend on the dark-matter content and on the relative rotation states of the constituent fluids. Throughout, rotation is treated as a first-order perturbation imposed on non-rotating equilibrium configurations, with background sequences constructed by varying the central densities of the nuclear and dark-matter components. This setup allows us to quantify how gravitational coupling between the fluids and the presence of a dark sector modify both the internal spacetime structure and global rotational observables, including the total moment of inertia and its role in equation-of-state-insensitive relations involving tidal deformability.

%%%%%%%%%%%%%%%%%%%%%%%%%%%%%
\subsection{Frame Dragging and Effective Moments of Inertia}
\label{sec:3a}
%%%%%%%%%%%%%%%%%%%%%%%%%%%%%

%%%%%%%%%%%%%%%%%%%%%%%%%%%%%
%  Figure 1
%%%%%%%%%%%%%%%%%%%%%%%%%%%%%
\begin{figure}[tbp]
    \centering
    \includegraphics[width=\columnwidth]{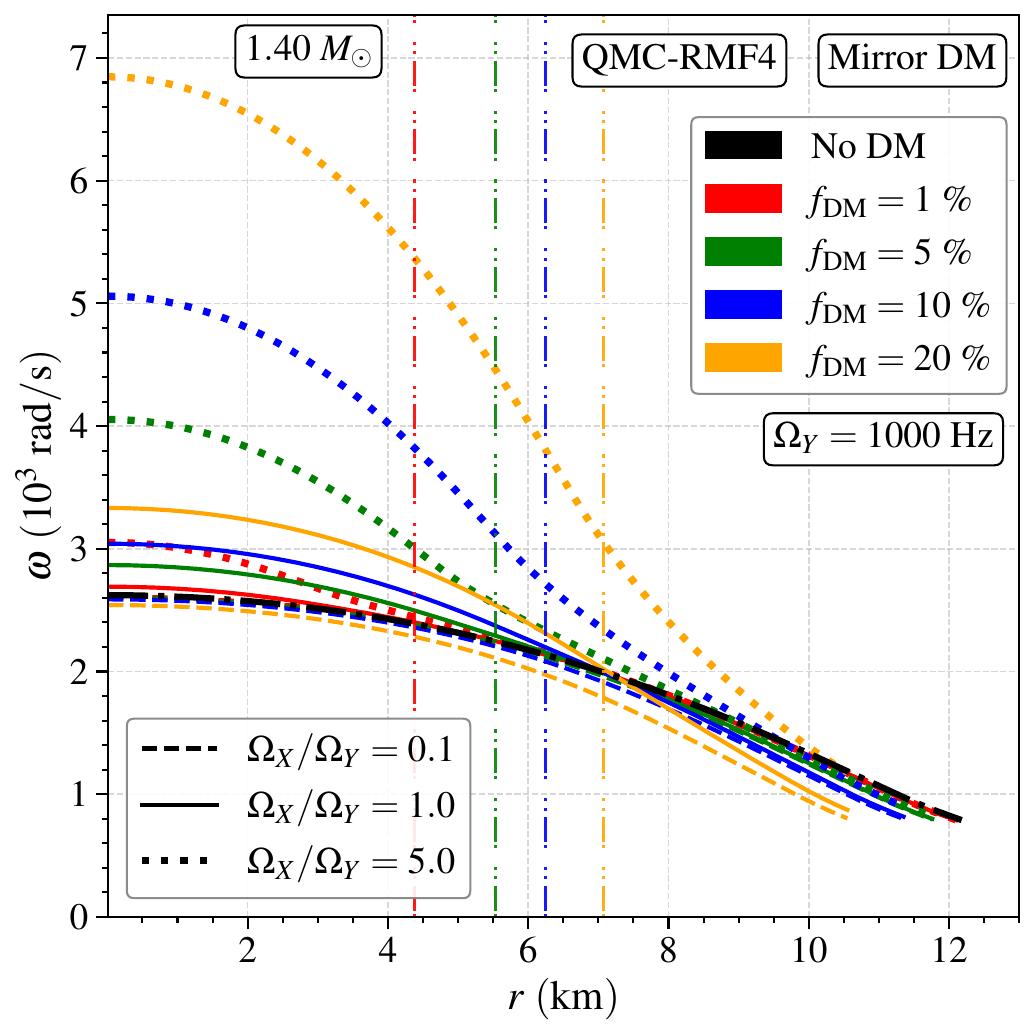}
    \caption{Radial profiles of the frame-dragging function $\omega(r)$ for a fixed-mass two-fluid neutron star with gravitational mass $M = 1.4\, M_\odot$, computed using the QMC-RMF4 nuclear equation of state in the mirror dark matter scenario. The outer fluid (fluid $Y$) is identified with the stellar surface and is assigned a fixed angular velocity $\Omega_Y = 1000$ Hz, while the inner fluid (fluid $X$) is allowed to rotate independently. Different colors correspond to different dark matter mass fractions $f_{\rm DM}$, ranging from the standard single-fluid neutron star configuration (without dark matter) described by the QMC-RMF4 equation of state (black) to two-fluid models with $f_{\rm DM}=1\%,\, 5\%,\, 10\%$, and $20\%$. Different line styles indicate the ratio of the rotation rates of the two fluids, $\Omega_X/\Omega_Y = 0.1$ (dashed), $1.0$ (solid), and $5.0$ (dotted). Vertical dot-dashed lines mark the surface radius of fluid $X$ for each dark matter fraction. The profiles illustrate how both the dark matter content and the relative rotation between the fluids modify the spacetime frame-dragging response throughout the stellar interior.}
    \label{fig:figure1}
\end{figure}
%%%%%%%%%%%%%%%

Figure~\ref{fig:figure1} shows radial profiles of the frame-dragging function $\omega(r)$ for neutron-star models with gravitational mass $M=1.40\, M_\odot$ computed using the QMC-RMF4 equation of state in the mirror dark-matter scenario. The outer fluid ($Y$) defines the stellar surface and is assigned a fixed angular velocity $\Omega_Y = 1000$ Hz, while the inner fluid ($X$) rotates at three representative relative rotation states given by the ratios $\Omega_X/\Omega_Y = 0.1,\ 1.0,\ 5.0$, corresponding to the dashed, solid, and dotted curves, respectively. The parameter $f_{\rm DM} = M_{\rm DM}/M_{\rm tot}$ specifies the fraction of the total gravitational mass that corresponds to dark matter at the level of the equilibrium solution, where $M_{\rm DM}$ denotes the gravitational mass associated with the dark-matter fluid and $M_{\rm tot}$ is the total gravitational mass of the configuration. The case $f_{\rm DM}=0$ corresponds to the standard single-fluid configuration shown by the black curve, whereas values $f_{\rm DM}=1\%,\, 5\%,\, 10\%,$ and $20\%$ represent two-fluid models and are shown with the corresponding colored (red, green, blue, and orange) curves. The interface between the dark-matter core (inner fluid $X$) and the surrounding nuclear layer (outer fluid $Y$) is indicated by vertical dot-dashed lines, whose outward shift with increasing $f_{\rm DM}$ reflects the growing radial extent of the dark-matter component.

Throughout this work, rotation is treated to first order in the Hartle–Thorne expansion, for which the frame-dragging equation is linear in the fluid angular velocities and admits the decomposition $\omega(r) = a_X(r)\, \Omega_X + a_Y(r)\, \Omega_Y$. Consequently, for a fixed non-rotating background the radial shape of $\omega(r)$ is invariant under a global rescaling of the spin frequencies, while its amplitude scales linearly with $\Omega_{X}$ and $\Omega_{Y}$; equivalently, $\omega(r)/\Omega_{Y}$ depends only on the equilibrium structure and on the ratio $\Omega_{X}/\Omega_{Y}$. The choice $\Omega_{Y} = 1000$ Hz in Fig.~\ref{fig:figure1} (and in Fig.~\ref{fig:J_pm_mirror}, where the ratios $J_{\pm}/J_T$ are evaluated for the co-rotating case $\Omega_X = \Omega_Y = 1000$ Hz) is therefore adopted solely as a convenient normalization. The formalism applies in the slow-rotation regime $\Omega \ll \Omega_{K}$, where $\Omega_{K}$ is the mass-shedding (Kepler) angular velocity.

Across all configurations shown in Fig.~\ref{fig:figure1}, the frame dragging function decreases monotonically with radius and asymptotically approaches a vacuum behavior proportional to $r^{-3}$ outside the star. For a fixed rotation-rate ratio $\Omega_X/\Omega_Y$, increasing the dark-matter fraction strengthens the influence of the inner fluid on the frame-dragging profile, leading to progressively larger deviations in $\omega(r)$ profile from the single-fluid reference throughout the stellar interior. As the dark-matter fraction $f_{\rm DM}$ grows at fixed total mass, the equilibrium pressure-energy profiles of both fluids adjust such that the relativistic enthalpy density $(\mathcal{E}_i + P_i)$ increases for each component in the inner region. Since this quantity multiplies $(\Omega_i - \omega)$ in the rotational driving term of Eq.~\eqref{eq:masterEq}, the contribution of the inner fluid to frame dragging becomes increasingly important. The sign of the resulting deviation--whether $\omega(r)$ is enhanced or suppressed relative to the single-fluid case--depends on the relative rotation of the inner component, a point made explicit below. As it approaches the star's surface, the profiles converge because the outer nuclear layer and the exterior matching condition jointly determine the asymptotic decay of $\omega(r)$ toward its vacuum behaviour.

The relative rotation of the two fluids modulates the interior structure of $\omega(r)$ in a qualitatively distinct manner. When the inner fluid rotates much more slowly than the outer layer ($\Omega_X/\Omega_Y = 0.1$), the frame-dragging profiles remain close to the standard single-fluid benchmark across all dark-matter fractions. In this regime, increasing $f_{\rm DM}$ shifts a larger portion of the total mass into the slowly rotating inner fluid, so that a greater share of the matter contributes through $(\mathcal{E}_X + P_X)$ but with a reduced rotational driving because $\Omega_X$ is small. As a result, the overall rotational response becomes slightly weaker than in the single-fluid configuration, causing the curves to lie marginally below the reference case at large $f_{\rm DM}$. In contrast, when the inner fluid rotates substantially faster than the outer layer ($\Omega_X/\Omega_Y = 5.0$), the rotational driving associated with the dark-matter core is strongly amplified, and the frame-dragging amplitude is noticeably enhanced in the inner region. This effect intensifies with increasing $f_{\rm DM}$ because a larger fraction of the stellar mass resides in the rapidly rotating component. The co-rotating case $\Omega_X/\Omega_Y = 1.0$ interpolates smoothly between these two limits. This interpolation is a direct consequence of the linear structure of the frame-dragging equation, which admits solutions of the form $\omega(r) = a_X(r)\, \Omega_X + a_Y(r)\, \Omega_Y$. For a fixed background configuration, the functions $a_X(r)$ and $a_Y(r)$ are fixed, and varying the ratio $\Omega_{X}/\Omega_{Y}$ continuously reweights the contributions of the two fluids to the total frame-dragging response. Consequently, the $\Omega_{X}/\Omega_{Y} = 1.0$ profiles fall between the suppressed ($\Omega_{X}/\Omega_{Y} = 0.1$) and enhanced ($\Omega_{X}/\Omega_{Y} = 5.0$) cases. Taken together, these trends show that the spatial structure of frame dragging is controlled jointly by the radial mass distribution between the two fluids and by how the total rotation is partitioned between them, with configurations featuring dense, rapidly rotating inner cores producing the strongest deviations from the single-fluid reference.
%\upd{We note that, for $\Omega_X/\Omega_Y=1$, the amplitude of $\omega$ in the inner region increases with $f_{\rm DM}$, while it conversely decreases in the outer region.}

Having established how the distribution of dark matter and the internal rotational partition between the two fluids shape the spatial profile of frame dragging, we now turn to global rotational diagnostics that quantify these effects in an integrated manner. In particular, we examine how embedding a dark-matter core modifies the total moment of inertia along equilibrium sequences as gravitational mass changes, thereby linking the local spacetime response discussed above to a quantity of direct astrophysical relevance.

%%%%%%%%%%%%%%%%%%%%%%%%%%%%%
%  Figure 2
%%%%%%%%%%%%%%%%%%%%%%%%%%%%%
\begin{figure}[tbp]
    \centering
    \includegraphics[width=\columnwidth]{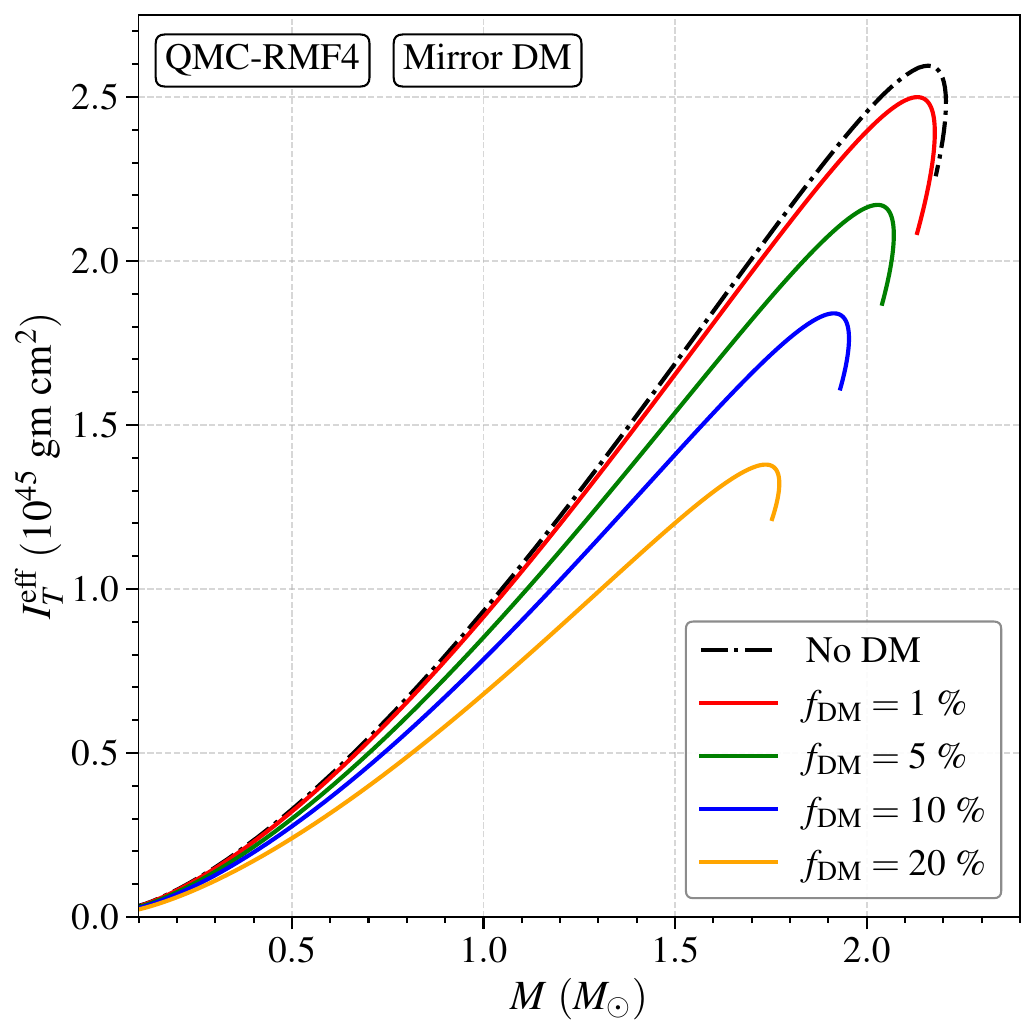}
    \caption{Effective total moment of inertia $I_{T}^{\rm{eff}}$ is plotted as a function of gravitational mass $M$ for neutron star models based on the QMC-RMF4 nuclear equation of state in the mirror dark matter scenario. The black dot-dashed curve corresponds to the standard single-fluid neutron stars described by the QMC-RMF4 equation of state (i.e., a configuration without dark matter), while the solid colored curves correspond to two-fluid configurations with increasing mirror dark-matter mass fraction $f_{\rm DM}=1\%,\, 5\%,\, 10\%$, and $20\%$, as indicated in the legend.}
    \label{fig:figure2}
\end{figure}
%%%%%%%%%%%%%%%
Figure~\ref{fig:figure2} shows the effective total moment of inertia $I_T^{\rm eff}$ as a function of gravitational mass $M$ for neutron-star models constructed with the QMC-RMF4 equation of state in the mirror dark-matter scenario. The black dot-dashed curve corresponds to the standard single-fluid configurations without dark matter, while the coloured curves display two-fluid sequences with fixed dark-matter mass fractions $f_{\rm DM}=1\%,\, 5\%,\, 10\%$, and $20\%$. For each value of $f_{\rm{DM}}$, equilibrium configurations are generated by varying the central densities of the nuclear and dark-matter fluids to span a continuous range of gravitational masses. The corresponding values of $I_T^{\rm eff}$ are computed from the inertia coefficients $I_{ij}$ introduced in Eq.~\eqref{eq:inertia_coeff}, and therefore depend solely on the underlying two-fluid background through the enthalpy density profiles and the frame-dragging basis functions $a_{X}(r)$ and $a_{Y}(r)$. Since the mirror dark-matter component obeys the same microphysical equation of state as the nuclear fluid, the differences between the curves arise entirely from how the total mass is repartitioned between the two fluids and from the way this redistribution modifies the gravitationally mediated rotational coupling, which is encoded in coefficients $I_{ij}$.

Across all sequences in Fig.~\ref{fig:figure2}, increasing the dark-matter fraction reduces $I_T^{\rm eff}$ at fixed gravitational mass, with the suppression becoming more pronounced for higher-mass stars. Since $I_T^{\rm eff}$ depends solely on the non-rotating background properties through the inertia coefficients $I_{ij}$ --~which in turn depend on the frame-dragging basis functions $a_X(r)$ and $a_Y(r)$~-- the behavior in Fig.~\ref{fig:figure2} reflects how the background mass distribution reshapes these basis functions rather than any change in the imposed rotation rates. In the equal-co-rotation case $\Omega_X = \Omega_Y \equiv \Omega$, i.e., $\Omega_X/\Omega_Y=1.0$, the normalized frame-dragging profile satisfies $\omega(r)/\Omega = a_X(r) + a_Y(r)$, so any change in the sum of the basis functions directly reflects how the two-fluid background modifies the spacetime response relative to the single-fluid benchmark. As $f_{\rm DM}$ increases, a larger share of the total mass resides at smaller radii within the dark-matter core. This inward redistribution of mass modifies the radial profiles of $a_X(r)$ and $a_Y(r)$ so that their sum is enhanced in the inner region but reduced in the outer layers, where the geometric lever arm $r^{4}$ most strongly weights the integrals defining $I_{ij}$. Although the increased inner enthalpy density locally amplifies frame dragging--visible in Fig.~\ref{fig:figure1} as an enhancement of $\omega(r)$ near the center--this contribution is evaluated at small radii where the factor $r^{4}$ is relatively small, and therefore contributes less to $I_{ij}$ than changes in the outer layers. Near the surface, where the $r^{4}$ weighting dominates, the frame-dragging amplitude systematically lies below the single-fluid reference, and this attenuation dominates the evaluation of $I_T^{\rm eff}$. The suppression is more substantial at larger gravitational masses because the configurations are more compact, further concentrating mass toward smaller radii and strengthening these geometric effects (attenuation of $a_{X}(r)$ and $a_{Y}(r)$ in the outer region). Thus, the ordering of curves with $f_{\rm DM}$ encapsulates how gravitationally mediated rotational coupling--encoded in the background-dependent basis functions--responds to radial mass redistribution: the local enhancement of frame dragging in the inner core does not offset the reduced lever arm at larger radii, leading to a net decrease of the integrated moment of inertia.

\begin{figure}[tbp]
    \centering
    \includegraphics[width=\columnwidth]{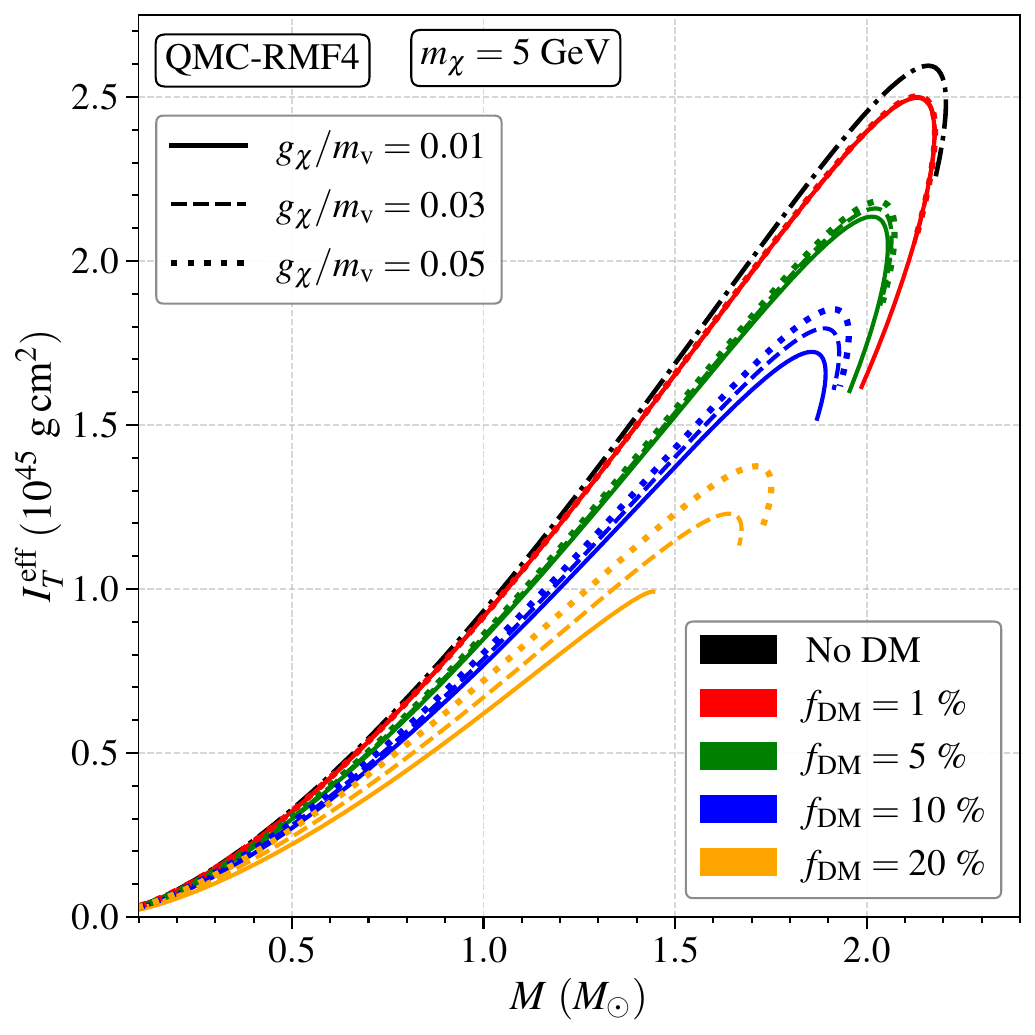}
    \caption{Effective total moment of inertia $I_{T}^{\rm eff}$ as a function of gravitational mass $M$ for two-fluid neutron stars with vector-interacting dark matter, computed using the QMC-RMF4 nuclear equation of state. The dark sector is characterized by $m_\chi = 5\ {\rm GeV}$ and coupling ratios $g_\chi/m_v = 0.01,\ 0.03,$ and $0.05$ MeV$^{-1}$ (solid, dashed, and dotted curves, respectively). Coloured curves correspond to dark-matter mass fractions $f_{\rm DM}=1\%,\ 5\%,\ 10\%,$ and $20\%$ (red, green, blue, and orange), while the black curve denotes the single-fluid reference without dark matter.}
    \label{fig:figure3}
\end{figure}

To assess how these geometric and background-dependent mechanisms respond when the dark sector microphysics is modified, we next examine the effective total moment of inertia in the case of vector-interacting dark matter, where the stiffness of the dark matter equation of state is controlled by the coupling ratio $g_{\chi}/m_{\rm v}$ (and the mass of dark matter particle $m_\chi$). Across all sequences in Fig.~\ref{fig:figure3}, the qualitative impact of increasing the dark matter fraction is similar to the mirror scenario: adopting $m_\chi=5$ GeV, for any fixed choice of $g_{\chi}/m_{\rm v}$, the effective total moment of inertia $I_{T}^{\rm eff}$ decreases as $f_{\rm DM}$ grows, and the suppression becomes more pronounced at higher gravitational masses. However, unlike the mirror case, the ordering of the curves at fixed $f_{\rm DM}$ now depends on the dark matter microphysics. For a fixed mass fraction, decreasing the coupling ratio $g_{\chi}/m_{\rm v}$ --- which yields a softer dark matter equation of state in our setup --- allows the dark component to reach higher inner densities while occupying a smaller radial volume~\cite{jksq-q6ty}. This concentrates a larger amount of mass deeper inside the star, steepening the decline of the magnitudes of $a_X(r)$ and $a_Y(r)$ in the outer layers, where the geometric lever arm $r^{4}$ most strongly weights the integrals defining $I_{ij}$. Consequently, the inner core exhibits locally enhanced frame dragging compared to the single-fluid reference. Because the dark matter region occupies a smaller radial extent, this enhancement contributes only weakly to $I_{T}^{\rm eff}$, while the stronger attenuation of $a_X(r)$ and $a_Y(r)$ in the outer layers--where the geometric lever arm $r^{4}$ is largest--dominates the net suppression. As a result, softer dark-matter models (smaller $g_{\chi}/m_{\rm v}$) systematically produce lower values of $I_{T}^{\rm eff}$ at fixed gravitational mass than their stiffer counterparts, demonstrating how dark sector microphysics modulates the background-dependent basis functions and thereby the integrated rotational response.

%%%%%%%%%%%%%%%%%%%%%%%%%%%%%
%  Figure 3
%%%%%%%%%%%%%%%%%%%%%%%%%%%%%
%%%%%%%%%%%%%%%

While $I_{T}^{\rm eff}$ characterizes the background-dependent rotational response independently of any specific choice of rotation rates, astrophysical observations are sensitive only to the rotation of the electromagnetically visible (nuclear) component of the star. In practice, pulsar timing and relativistic spin-orbit measurements do not provide a direct estimate of the total angular momentum $J_{T}$, but rather constrain an effective moment of inertia associated with the observed spin frequency of the nuclear surface. Within the two-fluid framework, the appropriate quantity that corresponds to such observational inferences is therefore, $I_{\rm obs} \equiv J_{T}/\Omega_{\rm{NM}}$, which expresses the total angular momentum normalized by the rotation rate of the nuclear matter fluid \footnote{Throughout this work, ‘observed moment of inertia’ refers to the effective inertia that would be inferred from rotational observables under the assumption of a single rigidly rotating star, rather than a direct measurement of the total angular momentum.}. In configurations where nuclear matter extends to the stellar surface (dark-matter core configurations), one has $\Omega_{\rm{NM}} = \Omega_{Y}$, while in dark-matter halo configurations the observed spin still corresponds to the nuclear component, $\Omega_{\rm{NM}} = \Omega_{X}$. In the co-rotating limit $(\Omega_{X} = \Omega_{Y})$, $I_{\rm obs}$ reduces to $I_{T}^{\rm eff}$ and therefore depends solely on the equilibrium stellar structure, while for non-co-rotating configurations $(\Omega_{X} \neq \Omega_{Y})$ it retains an explicit dependence on the internal rotation ratio $\Omega_{X}/\Omega_{Y}$, thereby encoding how the unobserved internal dynamics influence the integrated rotational response measured at the stellar surface. Two-dimensional contour plots of $I_{\rm obs}$ across the central energy densities parameter space $(\mathcal{E}_{c}^{\rm NM},\, \mathcal{E}_{c}^{\rm DM})$ for both mirror and vector dark-matter models are provided in Appendix~\ref{sec:appendx1}.

While $I_{\rm obs}$ provides the correct bridge to observations when defined with respect to the nuclear rotation rate, caution is still required when interpreting its value in dark-matter halo configurations $(R_{\rm DM} > R_{\rm NM})$. In such cases, part of the total angular momentum $J_{T}$ is carried by material external to the baryonic surface, and its contribution is not directly tied to the observed spin frequency. Consequently, although $I_{\rm obs} \equiv J_{T}/\Omega_{\rm{NM}}$ remains the quantity relevant for observational inference, it should be understood as an effective inertia that encodes both directly observed and gravitationally coupled components. Only in core-type configurations--where the nuclear layer reaches the stellar surface--does $I_{\rm obs}$ coincide straightforwardly with the moment of inertia that would be inferred under a single-fluid assumption.

%%%%%%%%%%%%%%%%%%%%%%%%%%%%%
\subsection{Intrinsic Rotational Modes and Eigenvalues of the Inertia Matrix}
\label{sec:3b}
%%%%%%%%%%%%%%%%%%%%%%%%%%%%%

A complementary perspective on the rotational response of two-fluid configurations is obtained by examining the eigen-moments of inertia $I_{+}$ and $I_{-}$, which characterize the intrinsic rotational degrees of freedom of the coupled system independently of the specific rotation rates imposed on the two fluids. Unlike the effective moments $I_{X}^{\rm{eff}}$ and $I_{Y}^{\rm{eff}}$, which measure how the total angular momentum responds to externally prescribed rotation rates $(\Omega_{X},\, \Omega_{Y})$, the eigenvalues $I_{\pm}$ are properties of the equilibrium background alone and quantify how efficiently the star generates angular momentum when it is excited in one of its two collective rotational modes. These modes arise from gravitational coupling between the fluids: neither corresponds to the rotation of an individual component, but instead each represents a coherent rotational pattern of the coupled configuration.

%%%%%%%%%%%%%%%%%%%%%%%%%%%%%
%  Figure 4
%%%%%%%%%%%%%%%%%%%%%%%%%%%%%
\begin{figure}[tbp]
    \centering
    \includegraphics[width=\columnwidth]{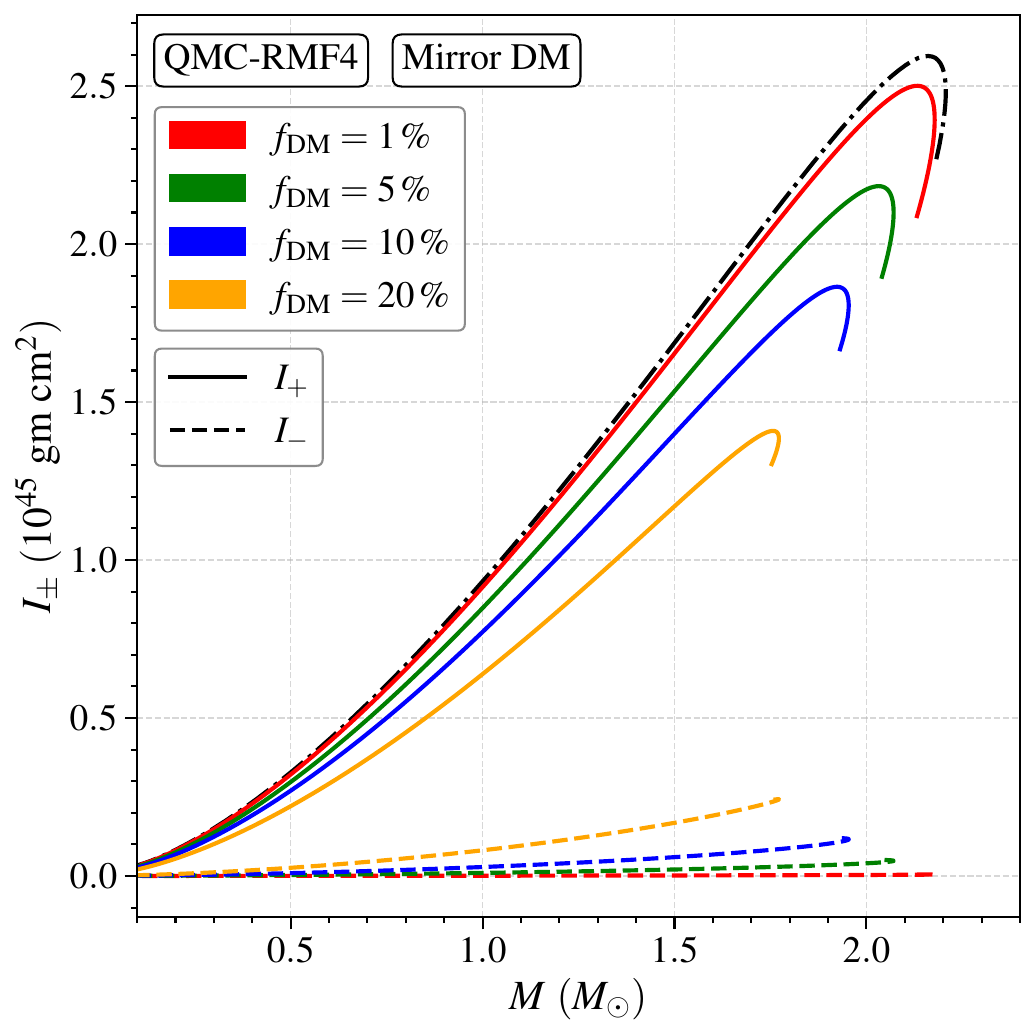}
    \caption{Eigen-moments of inertia $I_{+}$ and $I_{-}$ as functions of the gravitational mass $M$ for two-fluid neutron stars in the mirror dark matter scenario, computed using the QMC-RMF4 equation of state. Colored curves correspond to increasing mirror dark matter mass fractions $f_{\rm{DM}} = 1\%,\, 5\%,\, 10\%$, and $20\%$ as indicated in the legend, while the two eigen-branches $I_{+}$ (solid) and $I_{-}$ (dashed) represent the intrinsic rotational response modes of the gravitationally coupled two-fluid configuration. The black dash-dotted curve corresponds to the standard single-fluid neutron star without dark matter described by the QMC-RMF4 equation of state.}
    \label{fig:I_pm_mirror}
\end{figure}
%%%%%%%%%%%%%%%
Figure~\ref{fig:I_pm_mirror} displays the eigen-moments of inertia $I_{+}$ (solid curves) and $I_{-}$ (dashed curves) as functions of gravitational mass $M$ for sequences of mirror dark matter models constructed with the QMC-RMF4 equation of state. Each coloured set of curves corresponds to a fixed dark matter mass fraction $f_{\rm{DM}}$, while the black dash-dotted curve denotes the single-fluid reference without dark matter. For all mass fractions considered, the two eigen-branches are cleanly separated: $I_{+}$ forms the dominant branch with the larger magnitude, whereas $I_{-}$ remains significantly smaller across the full mass range. This behaviour reflects the structure of the inertia matrix, where the diagonal terms associated with each fluid are comparable while the off-diagonal cross-couplings remain subdominant. As a result, the collective mode associated with $I_{+}$ couples more efficiently to the total mass distribution and therefore carries the bulk of the intrinsic rotational inertia, while $I_{-}$ corresponds to a weaker, counterbalancing rotational channel in which the coupled response of the two components partially cancels.

The ordering and evolution of the curves with increasing $f_{\rm{DM}}$ encode how the internal mass distribution reshapes these intrinsic rotational channels. For both eigen-branches, the introduction of even a modest dark-matter core ($f_{\rm{DM}} \sim 1\%$) reduces the magnitude of the eigen-moments relative to the single-fluid sequence, with the suppression becoming progressively larger as the dark matter fraction grows. The physical origin of these trends is closely related to the geometric mechanism identified earlier for $I_{T}^{\rm{eff}}$: transferring mass inward into a centrally concentrated dark component steepens the radial mass profile and diminishes the contribution from outer radii, where the geometric lever arm is largest. For the dominant branch $I_{+}$, whose eigen-rotation pattern is weighted more strongly toward the bulk, large-radius motion, this leads to a systematic suppression with increasing $f_{\rm{DM}}$, mirroring the behaviour of $I_{T}^{\rm{eff}}$. In contrast, the subdominant branch $I_{-}$ places relatively more weight on the difference in redistribution of inertia between the inner dark core and the outer nuclear layer. As the dark component becomes more centrally concentrated, the increased gravitational leverage of the inner mass enhances this differential contribution, causing $I_{-}$ to exhibit a mild rise at higher $f_{\rm{DM}}$ and toward the high-mass end of the sequence. The opposite response of the two branches causes the gap between $I_{+}$ and $I_{-}$ to shrink as the dark-matter fraction grows, consistent with the behaviour seen in Fig.~\ref{fig:I_pm_mirror}.

The ratio $I_{-}/I_{+}$ thus offers a compact diagnostic of how strongly the secondary collective mode participates in the rotational response. Although we do not plot this ratio separately here, its behaviour can be inferred directly from the relative vertical spacing between the two branches: for lower dark matter fractions $(f_{\rm{DM}} \sim 1\,\%)$, the separation between $I_{+}$ and $I_{-}$ is large, meaning that the rotational dynamics are dominated by the primary mode. As $f_{\rm{DM}}$ increases, the two branches move closer together in fractional terms, indicating that the secondary mode becomes increasingly relevant to the rotational response, even though its absolute magnitude remains much smaller than that of $I_{+}$. Physically, this reflects the fact that as dark matter occupies a larger share of the inner volume, its increased gravitational weight and coupling to the nuclear component enhances the influence of inner regions on the subdominant mode. The result is that the subdominant rotational channel ($I_{-}$) captures a progressively more noticeable fraction of the intrinsic rotational inertia, even though it always remains smaller than $I_{+}$. In summary, increasing the dark matter fraction does not overturn the hierarchy of rotational modes, but it rebalances it: $I_{+}$ continues to dominate the intrinsic rotational response, yet $I_{-}$ becomes less negligible, reflecting the growing dynamical influence of mass concentrated in the inner regions.

Taken together, the behaviour of $I_{+}$ and $I_{-}$ demonstrates that embedding dark matter modifies not only global rotational observables such as $I_{T}^{\rm{eff}}$, but also the internal structure of the rotational response encoded in the eigen-moments. The dominant branch $I_{+}$ tracks the overall rotational capacity of the configuration and decreases systematically with increasing dark matter content, while $I_{-}$ carries a smaller but progressively more non-negligible share of the intrinsic inertia as the dark component grows. This decomposition provides a natural basis for examining how different intrinsic rotational modes contribute to the total angular momentum. We now further pursue this point by analysing the fractional mode contributions to $J_{T}$ under equal rotation of the two fluids.

We have verified that the qualitative structure of the eigen-moment branches persists when the dark sector is modeled using a vector-mediated fermionic equation of state: for $m_{\chi} = 5$ GeV and the full range of couplings $g_{\chi}/m_{\rm v} = 0.01,\, 0.03$, and $0.05$ MeV$^{-1}$, the ordering, separation, and systematic trends of $\{I_{+},\ I_{-}\}$ follow the same pattern as in the mirror case, indicating that the redistribution of mass and stiffness of the dark component reshapes the intrinsic rotational channels in a manner that is robust across the dark-matter microphysics considered here.

%%%%%%%%%%%%%%%%%%%%%%%%%%%%%
%  Figure 5
%%%%%%%%%%%%%%%%%%%%%%%%%%%%%
\begin{figure}[tbp]
    \centering
    \includegraphics[width=\columnwidth]{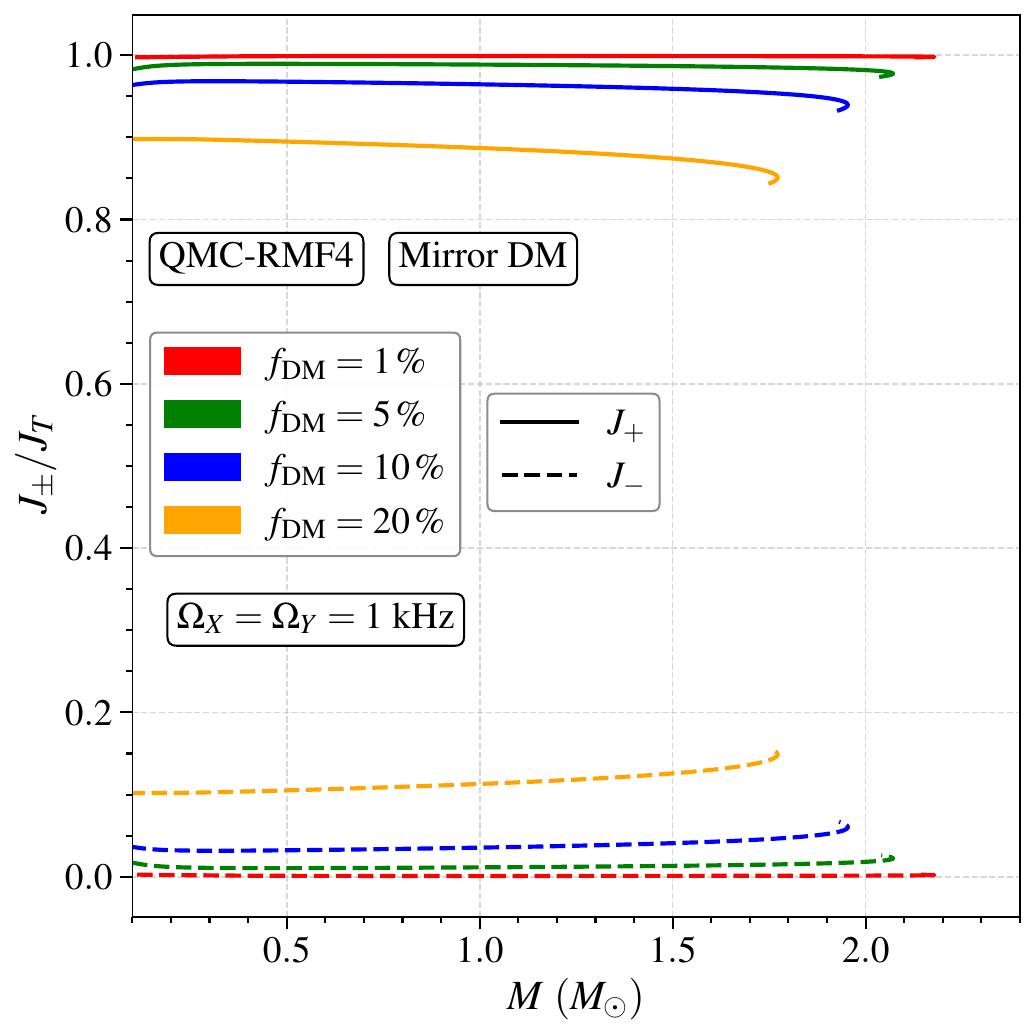}
    \caption{Fractional angular-momentum contributions of the two intrinsic rotational modes as functions of gravitational mass $M$ for mirror dark matter admixtures using QMC-RMF4 equation of state. Shown are the ratios $J_{+}/J_{T}$ (solid) and $J_{-}/J_{T}$ (dashed), evaluated for co-rotating configurations with $\Omega_{X} = \Omega_{Y} = 1000$ Hz. Colored curves correspond to increasing dark-matter mass fractions $f=1\%,\, 5\%,\, 10\%$, and $20\%$ (red, green, blue, orange).}
    \label{fig:J_pm_mirror}
\end{figure}
%%%%%%%%%%%%%%%

A complementary diagnostic of how the intrinsic rotational modes participate in the total spin budget is provided in Fig.~\ref{fig:J_pm_mirror}, which shows the fractional angular momenta $J_{+}/J_{T}$ and $J_{-}/J_{T}$ for co-rotating configurations with $\Omega_{X} = \Omega_{Y} = 1000$ Hz (Here, $J_{\pm} = I_{\pm}\Omega_{\pm}$ are the collective rotation rates obtained by projecting the fluid angular-velocity vector $(\Omega_{X},\, \Omega_{Y})$ onto the eigenbasis that diagonalizes the inertia matrix). Whereas the eigen-moments of inertia $\{I_{+},\, I_{-}\}$ characterize the capacity of each intrinsic mode to store angular momentum, the ratios $J_{\pm}/J_{T}$ reveal how much of the actual angular momentum of a corotating system is carried by each mode~\footnote{Because the intrinsic angular velocities $\Omega_{\pm}$ scale linearly with the imposed fluid rotation rate under co-rotation, the mode-participation ratios $J_{\pm}/J_{T}$ are invariant under changes in the absolute rotation frequency.}. In this sense, the curves should be interpreted as mode-participation factors evaluated along realistic sequences rather than as purely geometric properties of the equilibrium star.

%%%%%%%%%%%%%%%%%%%%%%%%%%%%%
%  Figure 6
%%%%%%%%%%%%%%%%%%%%%%%%%%%%%
\begin{figure*}[tbp]
    \centering
    \includegraphics[width=\textwidth]{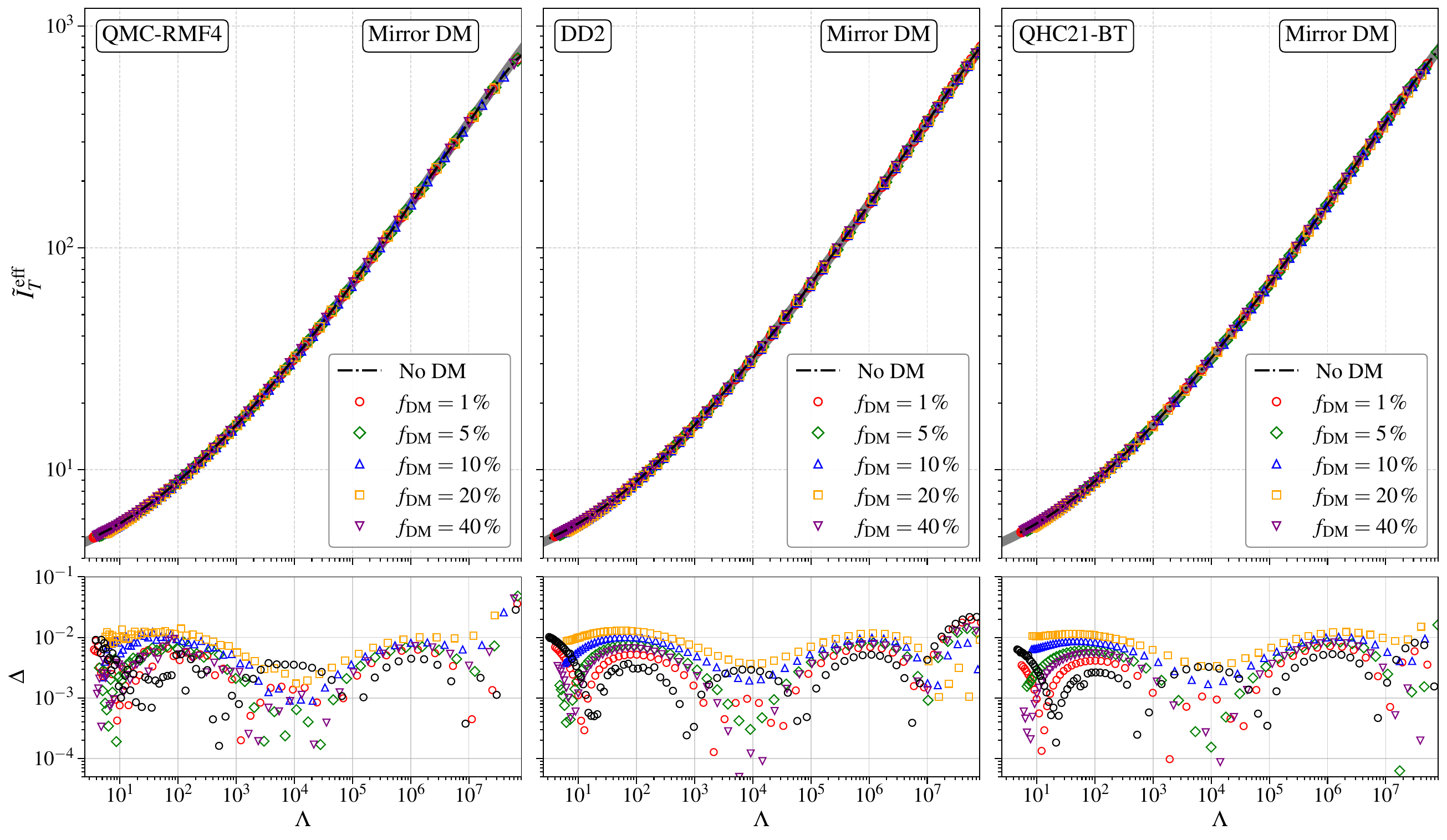}
    \caption{Universal relation between the dimensionless effective total moment of inertia, $\tilde I_{T}^{\rm eff} \equiv I_{T}^{\rm eff}/M^{3}$, and the tidal deformability $\Lambda$ for two-fluid neutron stars in the mirror dark matter scenario. The top panels show $\tilde I_{T}^{\rm eff}$ as a function of $\Lambda$ for three nuclear matter equations of state (from left to right: QMC-RMF4, DD2, and QHC21-BT). The dash-dotted black curve denotes the standard single-fluid neutron star sequence (no dark matter), while the colored symbols correspond to two-fluid configurations with mirror dark matter mass fractions $f_{\rm DM}=1\%,\,5\%,\,10\%,\,20\%,$ and $40\%$ (as indicated in the legend). The thick solid gray curve shows the baseline $\tilde I$--$\Lambda$ empirical relation obtained from the single-fluid analysis in Fig.~\ref{fig:ur-sf}. The bottom panels display the relative deviation $\Delta$ of the two-fluid neutron star data from this baseline relation. Across all three equations of state and over the full range of $\Lambda$ shown, the effective total moment of inertia $\tilde I_{T}^{\rm eff}$ preserves the $\tilde{I}-\Lambda$ universality to high accuracy despite the presence of a gravitationally coupled dark-matter component.}
    \label{fig:ur-mirror}
\end{figure*}
%%%%%%%%%%%%%%%

Across all dark matter fractions plotted in Fig.~\ref{fig:J_pm_mirror}, the primary mode associated with $I_+$ remains overwhelmingly responsible for the total spin: $J_{+}/J_{T} \gtrsim 0.85$ over most of the mass range, while $J_{-}/J_{T}$ contributes at the level of a few to $\lesssim 15 \%$. A systematic trend nevertheless emerges. As the concentration of the dark component increases, the enhanced inner weighting of the subdominant mode slightly raises $J_{-}/J_{T}$ -- most noticeably toward the high-mass end of each sequence, where the nuclear layer is also compact. The corresponding reduction in $J_{+}/J_{T}$ at fixed mass illustrates the same rebalance of intrinsic rotational channels already seen in the behavior of eigen-moments of inertia: the primary mode remains dominant, but the secondary mode becomes less negligible as the dark matter occupies a larger share of the inner volume. Importantly, this hierarchy persists even in the maximally symmetric mirror configurations, for example, at $f_{\rm DM} = 0.50$, where both fluids contribute equal mass and extend over the same radial domain.

It is worth emphasizing that equal microphysics (as in the case of mirror dark matter scenario) and equal rotation rates do not collapse the two collective channels into one: the off-diagonal inertia terms $I_{XY}$ and $I_{YX}$ generated by gravitational coupling ensure that the inertia matrix cannot be simultaneously diagonal in the fluid basis, and the configuration therefore retains two distinct intrinsic rotational modes. In other words, co-rotation of the fluids ($\Omega_{X} = \Omega_{Y}$) does not imply co-rotation of the intrinsic modes ($\Omega_{+} \neq \Omega_{-}$ in general); the latter are determined by diagonalising the coupled inertia tensor, not by prescribing independent fluid motion. The result is that the coupled two-fluid configuration necessarily supports two intrinsic collective modes whose angular-momentum participation differs, and this difference is sustained by the gravitational entanglement of the two components.

%%%%%%%%%%%%%%%%%%%%%%%%%%%%%
%  Figure 7
%%%%%%%%%%%%%%%%%%%%%%%%%%%%%
\begin{figure*}[tbp]
    \centering
    \includegraphics[width=\textwidth]{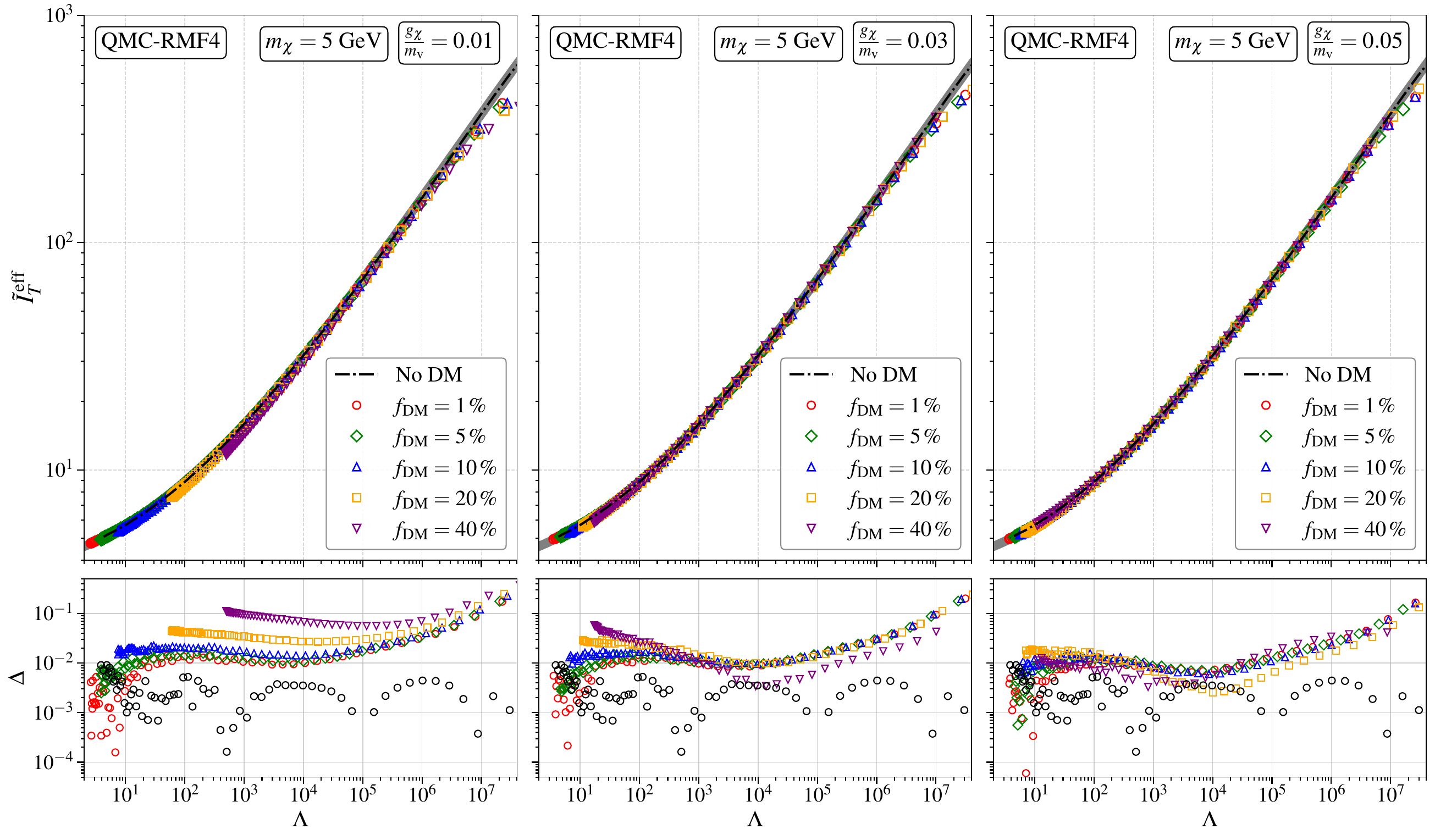}
    \caption{Universal relation between the dimensionless effective total moment of inertia, $\tilde I_{T}^{\rm eff} \equiv I_{T}^{\rm eff}/M^{3}$, and the tidal deformability $\Lambda$ for two-fluid neutron stars with QMC-RMF4 nuclear matter and fermionic vector dark matter with $m_{\chi} = 5$ GeV. The three columns correspond to different dark-sector coupling strengths $g_{\chi}/m_{\rm v} = 0.01,\, 0.03,$ and $0.05$ MeV$^{-1}$ (left to right). In the top panels, the dash-dotted black curve shows the standard single-fluid QMC-RMF4 sequence (no dark matter), colored symbols represent two-fluid configurations with dark-matter mass fractions $f_{\rm DM}=1\%,\,5\%,\,10\%,\,20\%,$ and $40\%$, and the thick solid gray curve denotes the baseline $\tilde{I}-\Lambda$ empirical fit obtained from the single-fluid analysis (Fig.~\ref{fig:ur-sf}). The bottom panels show the relative deviation $\Delta$ of the two-fluid effective total moment of inertia $I_{T}^{\rm{eff}}$ from the baseline single-fluid $\tilde{I}-\Lambda$ fit, with the same color and symbol scheme as in the top panels to identify deviations for the corresponding dark matter mass fractions.}
    \label{fig:ur-5g}
\end{figure*}
%%%%%%%%%%%%%%%

This mild but coherent reshaping of the spin budget carries several physical implications. The fact that even in co-rotation the secondary mode absorbs a nonzero and gradually increasing fraction of the angular momentum implies that perturbations capable of preferentially exciting this differential channel could access a reservoir of rotational inertia that is dynamically small but not negligible. In phenomenological terms, such a reservoir aligns with expectations from two-fluid neutron-star dynamics, where weakly coupled rotational responses may contribute to free-precession damping, glitch relaxation, or post-glitch mode excitation once superfluidity and entrainment are incorporated into the framework. We do not model such effects here, but the trends in Fig.~\ref{fig:J_pm_mirror} identify where that additional rotational “headroom’’ resides in the background structure. In this sense, the two-mode structure we quantify here provides the static background against which dynamical glitch models -- including superfluid entrainment and inter-component torque -- could be systematically developed. In this context, it is worth noting that recent observationally motivated studies of pulsar glitches and moment-of-inertia constraints in multi-component neutron stars (see Fig.~8 of Ref.~\cite{Basu2025}) implicitly probe an effective angular-momentum-carrying inertia, without resolving how this quantity decomposes into intrinsic collective modes of a coupled two-fluid system.

%%%%%%%%%%%%%%%%%%%%%%%%%%%%%
\subsection{Universal Rotational–Tidal Relations}
\label{sec:3c}
%%%%%%%%%%%%%%%%%%%%%%%%%%%%%

A final element of the rotational response that bridges microscopic physics and macroscopic observables is the well-established universality (I-Love relation) between the dimensionless moment of inertia $\tilde{I} \equiv I/M^{3}$ and the tidal deformability $\Lambda$. The dimensionless tidal deformability is defined in terms of the quadrupolar electric-type Love number $k_2$ as
%%%%%%%%%%%%%%%
\begin{align}
\Lambda \equiv \frac{2}{3} k_2 \left(\frac{R}{M}\right)^5 ,
\end{align}
%%%%%%%%%%%%%%%
where $R$ and $M$ denote the stellar radius and gravitational mass, respectively. The Love number $k_2$ characterizes the linear response of the star’s mass quadrupole moment $Q_{ij}$ to an external static tidal field $\mathcal{E}_{ij}$ through $Q_{ij} = - \lambda\, \mathcal{E}_{ij}$, with $\lambda \equiv \Lambda\, M^{5}$~\cite{Hinderer_2008, PhysRevD.107.115028, PhysRevD.81.123016}. In the single-fluid case, this I-Love relation provides a powerful connection, independent of the equations of state, between rotation and tidal response, enabling joint constraints on neutron-star structure from pulsar timing and gravitational-wave observations without detailed knowledge of the microphysics.

Before assessing how this behaviour extends to two-fluid configurations, it is useful to recall the degree to which the $\tilde{I}$-$\Lambda$ universality holds for standard neutron stars. Figure~\ref{fig:ur-sf} (Appendix~\ref{sec:appendx2}) displays $\tilde{I}(\Lambda)$ for a representative set of realistic equations of state spanning the stiffness range relevant to realistic stellar matter. The data collapse onto a single smooth curve with fractional deviations $\Delta \lesssim \mathcal{O}(10^{-2})$ across the relevant range of $\Lambda$, confirming that the empirical fit
%%%%%%%%%%%%%%%
\begin{align}
    \log_{10}\tilde{I} = \sum_{n=0}^{4} d_{n} \, (\log_{10}\Lambda)^{n}
\end{align}
%%%%%%%%%%%%%%%
captures the equation of state dependence to percent-level accuracy\footnote{The numerical values of the coefficients $d_n$ are listed in Appendix~\ref{sec:appendx2}.}. This single-fluid baseline establishes the reference for quantifying the influence of a gravitationally coupled dark-matter component.

%%%%%%%%%%%%%%%%%%%%%%%%%%%%%
%  Figure 8
%%%%%%%%%%%%%%%%%%%%%%%%%%%%%
\begin{figure*}[tbp]
    \centering
    \includegraphics[width=\textwidth]{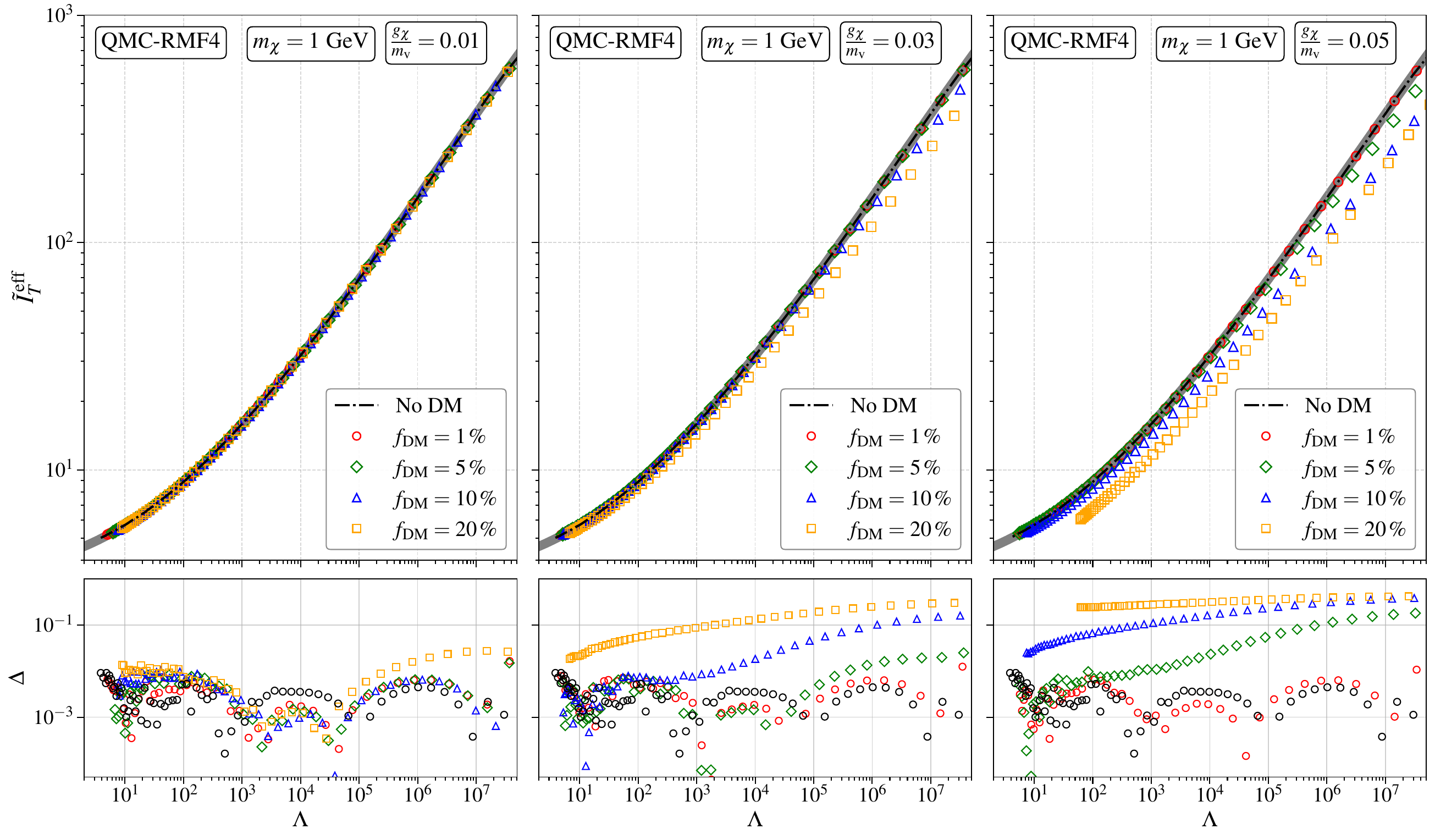}
    \caption{Same as Fig.~\ref{fig:ur-5g} but for $m_{\chi} = 1$ GeV}
    \label{fig:ur-1g}
\end{figure*}
%%%%%%%%%%%%%%%

Having established the single-fluid baseline, we now examine whether the presence of a gravitationally coupled dark component disrupts the $\tilde{I}-\Lambda$ mapping.
For this purpose, we focus on the effective total moment of inertia, $I_{T}^{\rm{eff}}$, defined from the total angular momentum of a corotating configuration. This quantity is the natural extension of the single-fluid moment of inertia to the two-fluid case as it measures the angular-momentum response of the whole star to a prescribed spin rate, irrespective of how the response is partitioned between the fluids or between intrinsic modes. By contrast, the intrinsic eigenvalues $I_{\pm}$, although rotational invariants constructed with the equilibrium background, characterize internal collective rotational channels rather than the global angular-momentum response of the configuration. Meanwhile, the observable moment $I_{\rm{obs}}$ depends explicitly on the chosen fluid rotation rates $(\Omega_{X},\, \Omega_{Y})$. For this reason, neither $I_{\pm}$ nor $I_{\rm{obs}}$ provides a unique, externally measurable generalization of the single-fluid moment of inertia appropriate for testing I-Love universality. Thus, $I_{T}^{\rm{eff}}$ is the appropriate quantity for testing whether global rotational-tidal correlations persist once an additional gravitationally bound fluid is introduced.

Figure \ref{fig:ur-mirror} displays $\tilde{I}_{T}^{\rm{eff}} \equiv I_{T}^{\rm{eff}}/M^{3}$ as a function of tidal deformability $\Lambda$ for three representative nuclear-matter equations of state (QMC-RMF4, DD2, QHC21-BT) across a wide range of mirror dark-matter mass fractions $f_{\rm{DM}}$. The top panels show that all two-fluid sequences lie nearly on top of the single-fluid reference curve. Despite variations in dark-matter content, stiffness of the nuclear equation of state, and global compactness, the data follow the same monotonic trend and exhibit no systematic drift away from the baseline. The bottom panels, which quantify the fractional deviation ($\Delta \equiv |\tilde{I}-\tilde{I}_{\rm fit}|/\tilde{I}$) relative to the single-fluid empirical fit, reveal that departures remain confined to the $\lesssim$ few-percent level for all three equations of state and for various dark matter mass fractions, over essentially the full range of $\Lambda$ shown. Notably, even for substantial dark-matter admixtures ($f_{\rm{DM}} \geq 20\%$), the deviations do not grow coherently with increasing compactness, indicating that the structural correlation between rotational response and tidal deformability remains robust.

However, the behaviour changes qualitatively when the dark sector possesses its own microphysical stiffness scale, as in the case of vector-mediated fermionic dark matter. Figures~\ref{fig:ur-5g} and~\ref{fig:ur-1g} show $\tilde{I}_{T}^{\rm{eff}} (\Lambda)$ for two-fluid stars constructed with QMC-RMF4 nuclear matter and vector dark matter at fixed particle masses $m_{\chi} = 5$ GeV and $m_{\chi} = 1$ GeV, respectively, for three representative couplings $g_{\chi}/m_{\rm{v}} = 0.01,\, 0.03$ and $0.05$ MeV$^{-1}$. In this model, the dark matter equation of state can become substantially softer or stiffer than the nuclear one depending on the combination of particle mass and coupling strength $(g_{\chi}/m_{\rm v})$: for fixed coupling the dark matter equation of state softens as $m_{\chi}$ increases, while for fixed $m_{\chi}$ it stiffens with increasing $g_{\chi}/m_{\rm v}$. Figures \ref{fig:ur-5g} and \ref{fig:ur-1g} demonstrate that these departures in stiffness reshape how pressure is contributed from and how mass is shared between the two fluids, thereby modifying the structural correlation between rotational inertia and tidal response. In all cases, as either the dark matter fraction or the stiffness contrast between the two components grows, systematic deviations from the single-fluid I-Love universal curve emerge, most visibly in cases where the dark component equation of state becomes significantly softer or stiffer than the nuclear one, for e.g., the model with $m_\chi=5$ GeV and $g_\chi/m_v=0.01$ MeV$^{-1}$ is the softest, while the model with $m_\chi=1$ GeV and $g_\chi/m_v=0.05$ MeV$^{-1}$ is the stiffest equation of state among the models shown in Figs.~\ref{fig:ur-5g} and~\ref{fig:ur-1g}.

In quantitative terms, the magnitude of the deviations closely tracks both the stiffness contrast between the two fluids and the dark matter mass fraction. When the dark sector differs only mildly from the nuclear equation of state -- for example, at $m_{\chi} = 1$ GeV with weak coupling $g_{\chi}/m_{\rm v} = 0.01$ MeV$^{-1}$, or at $m_{\chi} = 5$ GeV with strong coupling $g_{\chi}/m_{\rm v}=0.05$ MeV$^{-1}$ -- the $\tilde{I}_{T}^{\rm eff}(\Lambda)$ sequences follow the single-fluid I-Love baseline to within a few percent even for $f_{\rm DM}\sim 20\%$. As the dark matter equation of state becomes much softer than the nuclear one, as in the $m_{\chi} = 5$ GeV and $g_{\chi}/m_{\rm v} = 0.01$ MeV$^{-1}$ case, the deviations grow systematically: for modest dark matter mass fractions the curves begin to fall below the baseline at large $\Lambda$ (low compactness), and for $f_{\rm DM}=40\%$ the discrepancy reaches $\sim 10\%$ over most of the sequence and rises to $\sim 30\%$ in the higher $\Lambda$ (the lower compactness) configurations. Increasing the coupling ratio to $g_{\chi}/m_{\rm v} = 0.03$ or $0.05$ MeV$^{-1}$ stiffens the dark matter equation of state and partially counteracts this softness-driven mismatch, reducing the deviations but not eliminating their growth with $f_{\rm DM}$ at higher $\Lambda$. A complementary trend emerges when the dark sector becomes much stiffer than the nuclear matter, as in the $m_{\chi}=1$ GeV case with stronger couplings ($g_{\chi}/m_{\rm v}=0.03$ and $0.05$ MeV$^{-1}$): here the two-fluid sequences again lie increasingly below the single-fluid I-Love relation as $f_{\rm DM}$ grows, with deviations that continue to increase with $\Lambda$ and reach $\sim 40-50\%$ for $f_{\rm DM}=20\%$, particularly in the high-$\Lambda$ (low-compactness) regime. Notably, the breakdown is markedly asymmetric, with the largest departures occurring for the light, strongly self-interacting dark sector (e.g. $m_{\chi} = 1$ GeV with $g_{\chi}/m_{\rm v}=0.05$ MeV$^{-1}$), where the stiffness contrast is maximal and the two-fluid sequences deviate most strongly from the single-fluid curve, while the $m_{\chi} = 5$ GeV sequences exhibit comparatively smaller departures over the same range of dark matter fractions. Nevertheless, it should also be noted that the relation between $\tilde{I}_{T}^{\rm eff}$ and $\Lambda$ for light dark matter particle ($m_\chi=1$ GeV) agrees well with the single-fluid I-Love relation, if the dark matter mass fraction is relatively very small, such as $f_{\rm DM}=1\%$.

Taken together, these results demonstrate that the behaviour of I-Love universality in two-fluid neutron stars is controlled not by the mere presence of an additional component, but by how strongly its microphysical stiffness departs from that of nuclear matter. In the mirror dark matter scenario, where both fluids share identical microphysics, the effective total moment of inertia preserves the single-fluid I-Love correspondence with remarkable accuracy within the explored set of nuclear equations of state, even at large dark-matter fractions. By contrast, vector-interacting dark matter introduces an independent stiffness scale that can either soften or stiffen the dark sector relative to the nuclear one, and once this contrast becomes pronounced, the correlated response between rotational inertia and tidal deformability is progressively lost. The effective moment of inertia therefore emerges as a sensitive probe of dark-sector microphysics: its adherence to, or departure from, universality encodes whether the dark component integrates smoothly into the global stellar structure or instead reshapes it in a manner that decouples rotational and tidal responses. From an observational perspective, this distinction is critical -- while mirror-like dark components remain largely invisible to I-Love tests, strongly interacting or light fermionic dark matter can leave clear, model-dependent imprints that invalidate equation-of-state-insensitive inference schemes and demand a genuinely two-fluid interpretation \footnote{Other universal relations, such as the $\Lambda$-compactness mapping, are broken even earlier in two-fluid stars~\cite{Sotani2025}, indicating that the I-Love correspondence is comparatively resilient but not symmetry-protected.}.

We emphasize that the universality discussed here is assessed across a representative set of realistic nuclear equations of state and two distinct dark-sector models; extending the analysis to a broader class of baryonic equation of state and dark matter microphysics would provide an even more comprehensive mapping of the robustness limits of the rotational–tidal correlation.
%%%%%%%%%%%%%%%%%%%%%%%%%%%%%
\section{Summary}
\label{sec:4}
%%%%%%%%%%%%%%%%%%%%%%%%%%%%%
In this work, we have developed a fully relativistic and self-consistent framework to describe the rotational properties of gravitationally coupled two-fluid neutron stars within the slow-rotation approximation. Treating the two components as independently conserved perfect fluids interacting only through spacetime curvature, we derived the coupled Tolman-Oppenheimer-Volkoff equations for the equilibrium background and obtained the corresponding frame-dragging equation governing rotational perturbations. Exploiting the linearity of the slow-rotation problem, we introduced a basis-decomposition of the frame-dragging response that allows the rotational behaviour of any two-fluid configuration to be constructed from background-dependent response functions, independent of the imposed rotation rates.

%%%%%%%%%%%%%%%%%%%%%%%%%%%%%
%  Figure 9
%%%%%%%%%%%%%%%%%%%%%%%%%%%%%
\begin{figure*}[htbp]
    \centering
    \includegraphics[width=\textwidth]{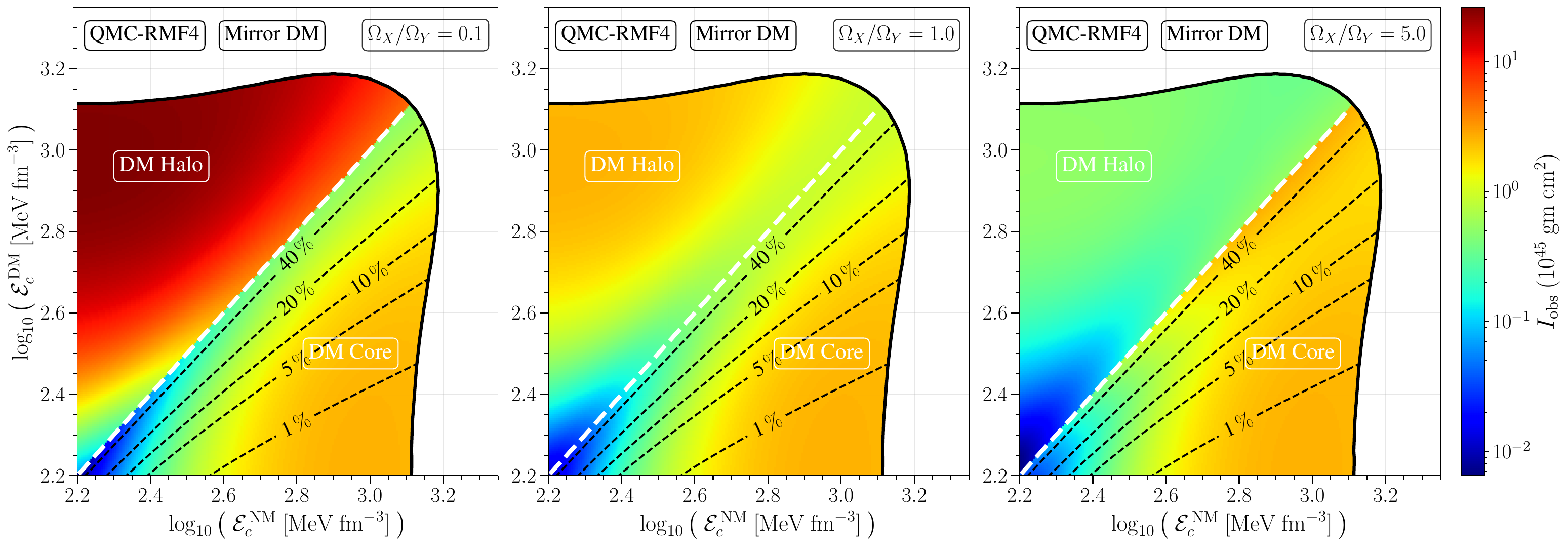}
    \caption{Observed moment of inertia $I_{\rm obs} \equiv J_{T}/\Omega_{\rm{NM}}$ for two-fluid neutron stars in the mirror dark matter scenario, computed using the QMC-RMF4 nuclear equation of state. In each panel, the color map represents the magnitude of $I_{\rm obs}$ (in units of $10^{45}\,\mathrm{g\,cm^{2}}$) over the two-dimensional parameter space spanned by the central energy densities of nuclear matter $({\cal E}_{c}^{\rm{NM}})$ and dark matter component $({\cal E}_{c}^{\rm{DM}})$ (both in $\mathrm{MeV\,fm^{-3}}$). From left to right, the three panels correspond to different relative rotation states of the two fluids, with $\Omega_{X}/\Omega_{Y} = 0.1$, $1.0$, and $5.0$, respectively. The thick solid black curve marks the stability boundary of two-fluid configurations~\cite{jksq-q6ty}, while the dashed black curves indicate contours of constant dark matter mass fraction $f_{\rm DM}=1\%,\,5\%,\,10\%$, and $20\%$. In the co-rotating case $\Omega_{X}=\Omega_{Y}$ (middle panel), the observed moment of inertia reduces to the effective total moment of inertia, $I_{\rm obs}=I_{T}^{\rm eff}$, which depends only on the equilibrium stellar structure and is independent of the rotation rate.}
    \label{fig:app1}
\end{figure*}
%%%%%%%%%%%%%%%

Building on this formulation, we established a transparent and physically meaningful definition of the total moment of inertia for two-fluid stars. The effective total moment of inertia, $I_{T}^{\rm{eff}}$, emerges naturally as a property of the equilibrium background and generalizes the single-fluid moment of inertia by consistently accounting for gravitationally mediated rotational coupling between the two components. In addition, by expressing the total angular momentum as a linear operator acting on the fluid rotation rates, we demonstrated that the rotational response of a two-fluid star can be decomposed into two intrinsic collective modes, characterized by eigen-moments of inertia $I_+$ and $I_-$. These eigenmodes represent fundamental rotational degrees of freedom of the coupled system and persist even in the limit of identical microphysics and co-rotation, highlighting that gravitational coupling alone is sufficient to sustain multiple intrinsic rotational channels.

Applying this framework to neutron stars containing either mirror dark matter or vector-interacting fermionic dark matter, we systematically explored how dark-matter content and microphysics reshape both local frame dragging and global rotational observables. We showed that increasing the dark-matter fraction redistributes mass toward smaller radii, leading to the deviation of frame dragging in the inner regions and a net suppression of the integrated moment of inertia due to the reduced geometric leverage at large radii. This geometric mechanism operates robustly across all equations of state considered and underlies the monotonic decrease of $I_{T}^{\rm{eff}}$ with increasing dark-matter fraction at fixed gravitational mass.

The intrinsic mode decomposition reveals additional structure beyond what is captured by global observables alone. While the dominant eigenmode $I_+$ continues to carry the majority of the rotational inertia, the subdominant mode $I_-$ becomes progressively more relevant as dark matter occupies a larger fraction of the inner volume. This redistribution reflects an increasing sensitivity of the rotational response to inner-core dynamics and identifies a small but non-negligible reservoir of rotational inertia associated with differential collective motion. Even in perfectly symmetric mirror configurations, gravitational coupling prevents the collapse of the rotational response into a single mode, emphasizing that two-fluid stars are intrinsically richer systems than their single-fluid counterparts.

A central result of this study concerns the robustness of the I-Love universality in the presence of a gravitationally coupled dark component. We demonstrated that when the dark sector shares the same microphysics as nuclear matter, as in the mirror dark matter scenario, the effective total moment of inertia preserves the single-fluid $\tilde{I}-\Lambda$ relation to percent-level accuracy across multiple nuclear equations of state and substantial dark-matter fractions. In this regime, the dark component integrates smoothly into the global structure, and the correlated response between rotation and tidal deformation remains intact. By contrast, when the dark sector possesses an independent stiffness scale, as in vector-mediated fermionic dark matter, departures from universality emerge once the stiffness contrast between the two fluids becomes pronounced. Much softer or stiffer dark matter significantly decouples rotational inertia from tidal deformability, leading to large, systematic deviations that invalidate equation-of-state-insensitive inference schemes based on single-fluid universality.

Taken together, our results demonstrate that the survival or breakdown of I-Love universality in two-fluid neutron stars is governed not by the mere presence of an additional component, but by how its microphysical stiffness compares to that of nuclear matter. The effective total moment of inertia, therefore, acts as a sensitive diagnostic of dark-sector microphysics, encoding whether the dark component participates coherently in the global structure or reshapes it in a manner that disrupts correlated rotational-tidal responses. More broadly, the formalism developed here provides a unified and physically transparent framework for interpreting rotational observables, intrinsic mode structure, and universal relations in multi-component relativistic stars, and establishes a solid foundation for future studies incorporating superfluidity, entrainment, dynamical glitches, and time-dependent rotational evolution within a genuinely two-fluid description.

%%%%%%%%%%%%%%%%%%%%%%%%%%%%%%%%%%%%%%%%%%%%%%%%
\acknowledgments
%%%%%%%%%%%%%%%%%%%%%%%%%%%%%%%%%%%%%%%%%%%%%%%%
This work is supported in part by the Japan Society for the Promotion of Science (JSPS) KAKENHI Grant Numbers 
JP23K20848  % Kiban(B) by Sotani
and JP24KF0090. % by Sotani & Kumar
%%%%%%%%%%%%%%%%%%%%%%%%%%%%%

%%%%%%%%%%%%%%%%%%%%%%%%%%%%%
\appendix
%%%%%%%%%%%%%%%
\section{Observed moment of inertia maps in the two-fluid parameter space}
\label{sec:appendx1}
%%%%%%%%%%%%%%%%%%%%%%%%%%%%%
In this appendix, we present global maps of the observed moment of inertia, $I_{\rm{obs}} \equiv J_{T}/\Omega_{\rm{NM}}$, across the two-fluid parameter space for both mirror and vector-interacting dark matter scenarios. Here $\Omega_{\rm NM}$ denotes the angular velocity of the nuclear (electromagnetically visible) component, which sets the observed spin frequency inferred from pulsar timing. These maps complement the one-dimensional co-rotating sequences ($\Omega_X=\Omega_Y$), where $I_{\rm obs}=I_T^{\rm eff}$,  shown in Figs.~\ref{fig:figure2} and~\ref{fig:figure3} by illustrating how $I_{\rm obs}$ varies over the full domain of stable two-fluid solutions in the plane of central energy densities plane of nuclear $({\cal E}_{c}^{\rm NM})$ and dark matter $({\cal E}_{c}^{\rm DM})$ components.

%%%%%%%%%%%%%%%%%%%%%%%%%%%%%
%  Figure 10
%%%%%%%%%%%%%%%%%%%%%%%%%%%%%
\begin{figure*}[tbp]
    \centering
    \includegraphics[width=\textwidth]{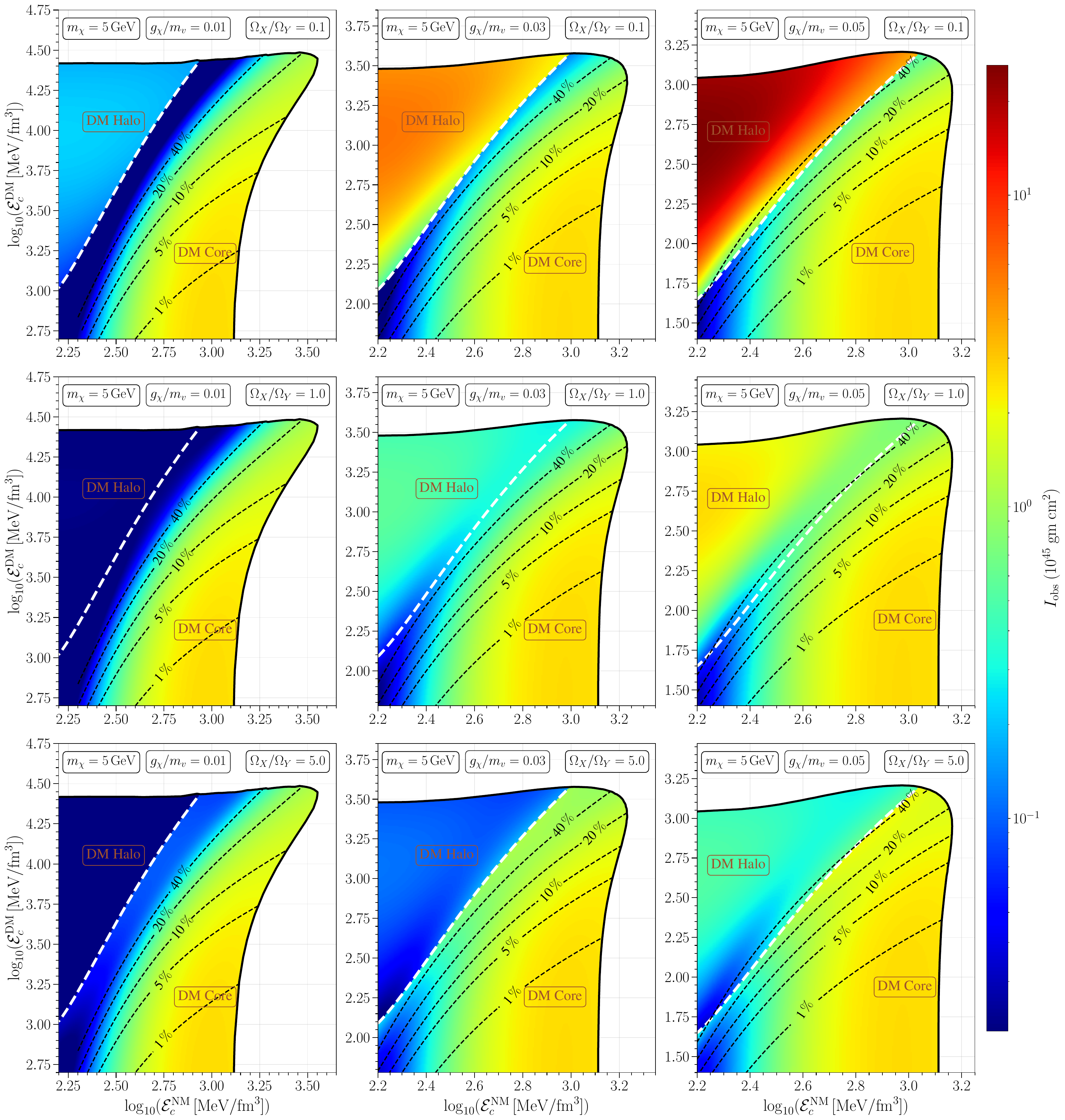}
    \caption{Observed moment of inertia $I_{\rm obs}\equiv J_T/\Omega_{\rm{NM}}$ for two-fluid neutron stars with the QMC-RMF4 as nuclear matter equation of state and a vector-mediated fermionic dark-matter component with particle mass $m_\chi=5~{\rm GeV}$. Each panel shows a color map of $I_{\rm obs}$ (in units of $10^{45}\,{\rm g\,cm^2}$) over the two-dimensional space of central energy densities, $(\log_{10}{\cal E}^{\rm NM}_c,\ \log_{10}{\cal E}^{\rm DM}_c)$, where ${\cal E}^{\rm NM}_c$ and ${\cal E}^{\rm DM}_c$ are in ${\rm MeV\,fm^{-3}}$. Columns correspond to different dark matter self-interaction strengths, $g_\chi/m_v=0.01,\ 0.03,$ and $0.05~{\rm MeV^{-1}}$ (left to right), while rows correspond to different relative rotation states of the two fluids, $\Omega_X/\Omega_Y=0.1,\ 1.0,$ and $5.0$ (top to bottom). The solid black curve marks the stability boundary of the two-fluid configurations. Black dashed curves indicate contours of constant dark matter mass fraction $f_{\rm DM}=1\%,\,5\%,\,10\%,\,20\%,$ and $40\%$. The white dashed curve in each panel separates dark-matter halo configurations ($R_{\rm DM}>R_{\rm NM}$) from dark-matter core configurations ($R_{\rm DM}<R_{\rm NM}$). In the co-rotating case $\Omega_X=\Omega_Y$ (middle row), the observed moment of inertia reduces to the effective total moment of inertia, $I_{\rm obs}=I_T^{\rm eff}$, which depends only on the equilibrium stellar structure and is independent of the rotation rate.}
    \label{fig:app2}
\end{figure*}
%%%%%%%%%%%%%%%

For the mirror dark matter case (QMC-RMF4 nuclear equation of state with a dark component described by the same equation of state), Fig.~\ref{fig:app1} shows colour contours of $I_{\rm obs}$ in the $({\cal E}_{c}^{\rm NM}, \, {\cal E}_{c}^{\rm DM})$ plane for three representative rotation ratios, $\Omega_X/\Omega_Y = 0.1,\, 1.0$, and $5.0$. The black solid curve delineates the onset of dynamical instability for static spherically symmetric stellar models, while the dashed curves trace constant dark matter mass fractions $f_{\rm DM}$. The white dashed line separates configurations with a dark matter core ($R_{\rm{DM}} < R_{\rm{NM}}$) from those with a dark matter halo ($R_{\rm{DM}} > R_{\rm{NM}}$). Because both fluids share the same equation of state, swapping their central densities and relabeling the components leaves $I_{\rm{obs}}$ invariant in the co-rotating case $(\Omega_{X} = \Omega_{Y})$, reflecting the interchange symmetry of the mirror two-fluid system.

At any given point in the $({\cal E}_{c}^{\rm NM},\, {\cal E}_{c}^{\rm DM})$ plane, the effective inertias $I_X^{\rm eff}$ and $I_Y^{\rm eff}$ are uniquely determined by the equilibrium background. Varying the rotation ratio $\Omega_X/\Omega_Y$ therefore does not alter the geometry of the stability boundary or the $f_{\rm{DM}}$ contours, but instead reweights the relative contributions of the two fluids to the observed inertia, depending on whether the nuclear component coincides with the inner or outer fluid.

In core-type configurations, where nuclear matter extends to the stellar surface, the observed rotation corresponds to $\Omega_{\rm{NM}} = \Omega_{Y}$, and
\begin{equation}
    I_{\rm obs} = I_Y^{\rm eff} + \left(\frac{\Omega_X}{\Omega_Y}\right) I_X^{\rm eff}.
    \label{eq:app_eq1}
\end{equation}
In this region, increasing $\Omega_{X}/\Omega_{Y}$ enhances the contribution of the inner (dark matter) fluid to $I_{\rm{obs}}$, producing progressively larger values as one moves from $\Omega_{X}/\Omega_{Y} = 1.0$ to $5.0$.

In halo-type configurations, where dark matter occupies the outer layers and nuclear matter is confined to the interior, the observed spin instead corresponds to $\Omega_{\rm{NM}} = \Omega_{X}$, so that
\begin{equation}
    I_{\rm obs} = I_X^{\rm eff} + \left(\frac{\Omega_Y}{\Omega_X}\right) I_Y^{\rm eff}.
    \label{eq:app_eq2}
\end{equation}
As a result, the same change in $\Omega_{X}/\Omega_{Y}$ produces the opposite effect in the halo region: for $\Omega_{X}/\Omega_{Y} = 0.1$, the factor $\Omega_{Y}/\Omega_{X}$ strongly amplifies the contribution of the outer dark matter layer, leading to the largest values of $I_{\rm{obs}}$, whereas for $\Omega_{X}/\Omega_{Y} = 5.0$ the contribution of the outer fluid is suppressed, yielding comparatively smaller $I_{\rm{obs}}$.

%%%%%%%%%%%%%%%%%%%%%%%%%%%%%
%  Figure 11
%%%%%%%%%%%%%%%%%%%%%%%%%%%%%
\begin{figure}[tbp]
    \centering
    \includegraphics[width=\columnwidth]{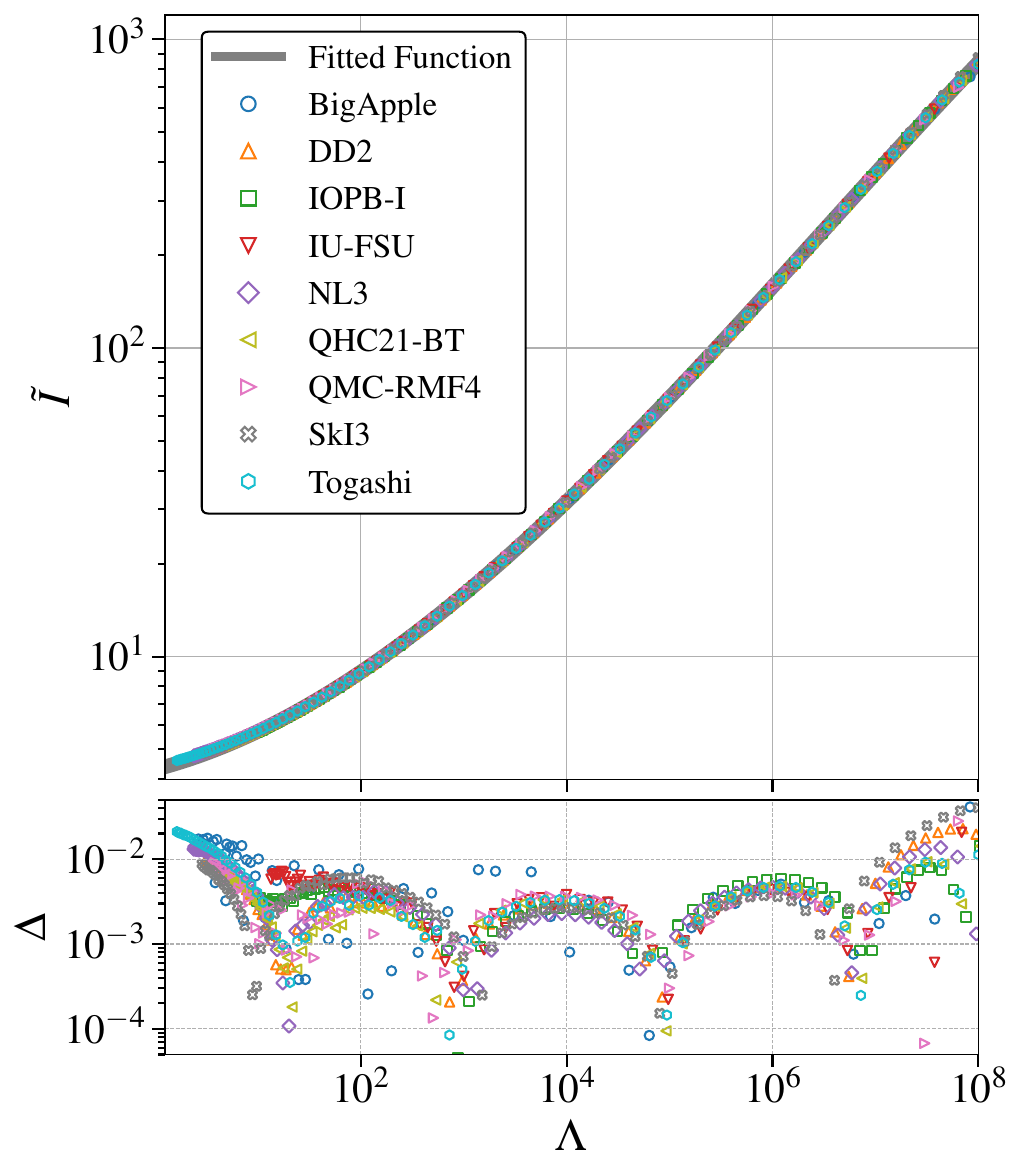}
    \caption{Universal relation between the (dimensionless) moment of inertia $\tilde{I}\, (\equiv I/M^{3})$ and the tidal deformability $\Lambda$ for standard single-fluid neutron stars. The upper panel shows $\tilde{I}(\Lambda)$ computed for a representative set of nuclear equations of state (BigApple, DD2, IOPB-I, IU-FSU, NL3, QHC21-BT, QMC-RMF4, SkI3, and Togashi), with the thick solid gray curve denoting the empirical fit obtained from the combined dataset. The lower panel displays the corresponding relative deviation $\Delta \equiv |\tilde{I}-\tilde{I}_{\rm{fit}}|/\tilde{I}$ as a function of $\Lambda$, quantifying the accuracy of the fitted relation across the full range of deformabilities shown.}
    \label{fig:ur-sf}
\end{figure}
%%%%%%%%%%%%%%%

This behaviour is clearly visible in Fig.~\ref{fig:app1}: above the core-halo boundary (in the halo region), the colour scale shifts to higher values for $\Omega_{X}/\Omega_{Y} = 0.1$ and to lower values for $\Omega_{X}/\Omega_{Y} = 5.0$, while the opposite trend holds in the core region. The mirror case, therefore, provides a particularly transparent illustration of how $I_{\rm{obs}}$ reflects purely kinematic reweighting of fixed background inertias, without introducing additional microphysical asymmetries between the two fluids.

Figure~\ref{fig:app2} shows the corresponding maps of the observed moment of inertia $I_{\rm{obs}}$ for vector-interacting dark matter with $m_{\chi} = 5$ GeV, now varying the interaction strength $g_{\chi}/m_{\rm{v}}$ across columns and the internal rotation ratio $\Omega_{X}/\Omega_{Y}$ down the rows. As before, the solid black curve marks the stability boundary of static two-fluid configurations, dashed black curves indicate constant dark matter mass fractions up to $f_{\rm DM} = 40\%$, and the white dashed line separates dark-core configurations $(R_{\rm{DM}} < R_{\rm{NM}})$ from dark-halo configurations $(R_{\rm{DM}} > R_{\rm{NM}})$.

Increasing the coupling strength $g_{\chi}/m_{\rm v}$ stiffens the dark matter equation of state and shifts the region of stable configurations toward lower dark matter central densities. In the color maps, this trend is accompanied by systematic shifts in the magnitude of $I_{\rm{obs}}$, most noticeably in halo-type configurations where a substantial fraction of the total angular momentum is carried by dark matter outside the baryonic surface. In the halo regime, the observed inertia is defined with respect to the nuclear rotation rate, $\Omega_{\rm{NM}} = \Omega_{X}$, so that $I_{\rm{obs}}$ follows the definition as in Eq.~\eqref{eq:app_eq2}. For a fixed coupling $g_{\chi}/m_{v}$ (i.e., fixed background), changing $\Omega_{X}/\Omega_{Y}$ therefore mainly reweights the contribution proportional to $I_{Y}^{\rm{eff}}$: the halo-region values in the top row $(\Omega_{X}/\Omega_{Y} = 0.1)$ are amplified by the factor $\Omega_{Y}/\Omega_{X}$, while in the bottom row $(\Omega_{X}/\Omega_{Y} = 5.0)$ this contribution is reduced. By contrast, the left-to-right variation across columns reflects a genuine microphysical effect: increasing $g_{\chi}/m_{v}$ stiffens the dark equation of state and shifts the stable domain toward lower ${\cal E}_{c}^{\rm{DM}}$, with a corresponding systematic increment in the magnitude of $I_{\rm{obs}}$ within the halo region.

%For fixed coupling strength, varying the rotation ratio $\Omega_{X}/\Omega_{Y}$ primarily redistributes the relative weight of the effective inertias $I_{X}^{\rm{eff}}$ and $I_{Y}^{\rm{eff}}$. Moving downward through each column (from $\Omega_{X}/\Omega_{Y} = 0.1$ to $5.0$) changes the normalization of $I_{\rm{obs}}$ across the halo region, while leaving the geometry of the stability boundary and the constant-$f_{\rm{DM}}$ contours unchanged. This confirms that the qualitative structure of the maps is set by the equilibrium background, whereas the overall magnitude of $I_{\rm{obs}}$ is controlled by how the total angular momentum is projected onto the nuclear rotation rate.

The halo-core division indicated by the white dashed line is therefore essential for interpreting these maps observationally. In core-type configurations, where nuclear matter defines the stellar surface, $I_{\rm obs}$ coincides directly with the moment of inertia inferred from pulsar timing observations. In halo-type configurations, the nuclear component is confined to an inner region while part of the total angular momentum is carried by dark matter outside the baryonic surface. In this case, $I_{\rm obs}$ should be interpreted as an effective inertia that encodes both the directly observed nuclear rotation and the gravitationally coupled contribution of the dark component.

Taken together, these global maps clarify how the interpretation of an observed moment of inertia depends on the internal composition and rotation state of a two-fluid neutron star. In particular, they identify the regions of parameter space in which timing-based measurements can be directly interpreted within a single-fluid framework, and those in which a two-fluid description is essential because part of the angular momentum is carried by a gravitationally coupled dark component outside the baryonic surface.

%%%%%%%%%%%%%%%%%%%%%%%%%%%%%
\section{Universality for single-fluid neutron stars}
\label{sec:appendx2}
%%%%%%%%%%%%%%%%%%%%%%%%%%%%%

Figure \ref{fig:ur-sf} summarizes the $\tilde{I}$-$\Lambda$ relation for standard single-fluid neutron stars, evaluated using a broad set of nuclear equations of state that span a wide range of nuclear stiffness. Both quantities are expressed in dimensionless form: $\tilde{I} \equiv I/M^{3}$ represents the normalized moment of inertia, while $\Lambda \propto k_{2}R^{5}/M^{5}$ is the dimensionless tidal deformability constructed from the quadrupolar Love number $k_{2}$ and the stellar radius $R$. The upper panel of Fig.~\ref{fig:ur-sf} shows that models based on BigApple~\cite{PhysRevC.102.065805}, DD2~\cite{PhysRevC.81.015803,PhysRevC.89.064321}, IOPB-I~\cite{PhysRevC.97.045806}, IU-FSU~\cite{PhysRevC.82.055803}, NL3~\cite{PhysRevC.55.540}, QHC21-BT~\cite{Kojo_2022}, QMC-RMF4~\cite{PhysRevC.106.055804}, SkI3~\cite{REINHARD1995467, PhysRevC.92.055803}, and Togashi~\cite{TOGASHI201778} all lie close to a single smooth curve despite their differing microphysics.

Fitting the combined dataset yields the empirical relation
%%%%%%%%%%%%%%%
\begin{align}
    \log_{10}\tilde{I} = \sum_{n=0}^{4} d_{n} \, (\log_{10}\Lambda)^{n} \nonumber
\end{align}
%%%%%%%%%%%%%%%
with coefficients $d_{0} = 0.633779$, $d_{1} = 0.085838$, $d_{2} = 0.038779$, $d_{3} = -0.001280$, and $d_{4} = - 0.000054$, shown by the thick solid grey line in the upper panel.\footnote{This polynomial was obtained from a least-squares fit over the full combined dataset of nuclear equations of state used here.} 

The lower panel displays the corresponding fractional deviation, $\Delta \equiv |\tilde{I}-\tilde{I}_{\rm fit}|/\tilde{I}$, demonstrating that the relation remains accurate to within a few percent across the entire range of $\Lambda$ encountered for realistic neutron-star models. These results define the single-fluid baseline against which two-fluid effects are assessed in Sec.~\ref{sec:3}.
%%%%%%%%%%%%%%%%%%%%%%%%%%%%%
% References
%%%%%%%%%%%%%%%%%%%%%%%%%%%%%
\bibliographystyle{apsrev4-2}
\bibliography{main.bib} % Replace with your .bib file
\end{document}